\documentclass[3p]{elsarticle}
\usepackage{lineno}
\usepackage{hyperref}
 \usepackage{tikz}
 \usepackage{setspace}
 \usepackage{subcaption}
 \usepackage{bm}
 \usepackage{color}
 \usepackage{tabularx,colortbl}
 \usepackage{pstricks-add}
\usepackage{graphicx,mwe}

\usetikzlibrary{arrows,shapes,automata,backgrounds,petri,positioning}
\usetikzlibrary{decorations.pathmorphing}
\usetikzlibrary{decorations.shapes}
\usetikzlibrary{decorations.text}
\usetikzlibrary{decorations.fractals}
\usetikzlibrary{decorations.footprints}
\usetikzlibrary{shadows}
\usetikzlibrary{calc}
\usetikzlibrary{spy}
\usetikzlibrary{calc,patterns,decorations.pathmorphing,decorations.markings}
\tikzstyle{spring}=[thick,decorate,decoration={zigzag,pre length=0.3cm,post length=0.3cm,segment length=6}]
\tikzstyle{damper}=[thick,decoration={markings,  
  mark connection node=dmp,
  mark=at position 0.5 with 
  {
    \node (dmp) [thick,inner sep=0pt,transform shape,rotate=-90,minimum width=15pt,minimum height=3pt,draw=none] {};
    \draw [thick] ($(dmp.north east)+(2pt,0)$) -- (dmp.south east) -- (dmp.south west) -- ($(dmp.north west)+(2pt,0)$);
    \draw [thick] ($(dmp.north)+(0,-5pt)$) -- ($(dmp.north)+(0,5pt)$);
  }
}, decorate]
\tikzstyle{process} = [rectangle, rounded corners, minimum width=3cm, minimum height=2.5cm, text 
centered, text width=3cm, draw=black, fill=red!30]
\tikzstyle{region} = [rectangle, minimum width=3.0cm, minimum height=2.5cm, 
text centered, draw=black, fill=white!30]
\tikzstyle{arrow} = [thick,->,>=stealth]

\title{A parallel dual-grid multiscale approach to CFD-DEM couplings}%

\author[rvt]{Gabriele Pozzetti\corref{cor1}}
\ead{gabriele.pozzetti@uni.lu}
\author[add1]{Hrvoje Jasak}
\author[rvt]{Xavier Besseron}
\author[rvt]{Alban Rousset}
\author[rvt]{Bernhard Peters}

\cortext[cor1]{Principal corresponding author. \textit{Email:}  
gabriele.pozzetti@uni.lu}
\address[rvt]{ Campus Belval, Université du Luxembourg
6, Avenue de la Fonte, L-4364 Esch-sur-Alzette  Luxembourg}
\address[add1]{ Faculty of Mechanical Engineering and Naval Architecture, University of Zagreb, Ivana
Lucica 5, Zagreb, Croatia}

\begin{document}

\begin{abstract}
In this work, a new parallel dual-grid multiscale approach for CFD-DEM couplings is investigated.
Dual-grid multiscale CFD-DEM couplings have been recently developed and successfully adopted in different applications still,
an efficient parallelization for such a numerical method represents an open issue.
Despite its ability to provide grid convergent solutions and more accurate results than standard CFD-DEM couplings,
this young numerical method requires good parallel performances in order to be applied to large-scale problems and, therefore, extend
its range of application.
The parallelization strategy here proposed aims to take advantage of the enhanced complexity of a
dual-grid coupling to gain more flexibility in the domain partitioning while keeping a low inter-process communication cost.
In particular, it allows avoiding inter-process communication between CFD and DEM software and still allows adopting 
complex partitioning strategies thanks to an optimized grid-based communication.
It is shown how the parallelized multiscale coupling holds all its natural advantages over a mono-scale coupling and 
can also have better parallel performance.
Three benchmark cases are presented to assess the accuracy and performance of the strategy.
It is shown how the proposed method allows maintaining good parallel performance when operated over 1000 processes.
\end{abstract}

\maketitle
\section{Introduction}
CFD--DEM couplings are nowadays widely used for the numerical description of engineering and technical problems featuring the interaction
between discrete entities and continuous fluid flows~\cite{Zhu20073378}.
In recent years, several attempts have been made to extend their applicability to large-scale scenarios and 
more and more complex configurations~\cite{BLAIS2016201, Wu2014347,NAG:NAG2387}.
The interaction between the discrete entities and the underlying fluid can be approached through direct numerical simulation~\cite{Uhlmann2005448,Mahmoudi20161091},
or via a so-called unresolved approach~\cite{doi:10.1021/i160024a007}. 
An unresolved or volume-averaged CFD-DEM coupling consists of the projection of Lagrangian fields into Eulerian ones 
by using volume-averaged variables\cite{doi:10.1021/i160024a007}.
In this kind of approaches, the coupling between CFD and DEM solution is performed
using cell values or locally interpolated values for the solution of the fluid-particle interactions~\cite{PozzettiIJMF}.
This significantly reduces the computational burden of the numerical scheme, and represents one of the main advantages of those approaches.
For this reason, unresolved CFD-DEM couplings are the most used for the solution of large-scale scenarios.

When coupling DEM with complex fluids, mono-scale volume averaged CFD-DEM couplings can have severe limitations~\cite{NAG:NAG2387}.
In particular, they can not always guarantee gird-convergent nor accurate solutions due to the fact that the CFD grids
adopted for the averaging cannot ensure optimal properties for the fluid solutions.
In~\cite{PozzettiIJMF}, a dual-grid multiscale approach was proposed in order to solve this problem. 
It consists of the identification of two length-scales within the particle-laden flow: a bulk scale at which the 
fluid-particle interaction is resolved, and a fluid fine-scale at which the equations governing the fluid 
phase are discretized.
Two grids are used in order to perform the two operations: a coarse grid for the bulk scale, and a fine grid 
for the fluid fine-scale. Eulerian fields are interpolated between the two grids in order to exchange 
information between the fine and the bulk scale.

As proposed in~\cite{PozzettiIJMF}, the dual-grid multiscale approach offers significant benefits over the mono-scale one 
in sequential execution. This approach has been shown 
to restore grid convergence of both local and averaged properties of the flow where the mono-scale approach cannot.
Additional works of~\cite{bpadditivemanufacturing}, ~\cite{POZZETTIPOWDERTEC,pozzettipowdermet} and ~\cite{WuLiang} 
relied on this method to tackle very 
complex engineering problems.
Nevertheless, in~\cite{PozzettiIJMF} we pointed out how the enhanced complexity of the multiscale scheme requires a 
thoughtful consideration for its parallel execution.
In particular, it was underlined how the rise in complexity of the multiscale approach with respect to 
a standard CFD-DEM coupling could represent an issue for its parallelization, as the introduced layer of inter-scale communication 
could potentially negatively affect the parallel performance of the software.

The parallelization of  CFD-DEM couplings is by itself a very difficult and open topic. First, the memory consumption 
of a coupled CFD-DEM simulation is normally important as, within a process, the information for both the CFD and DEM domain 
must be stored. Second, in addition to the intra-physic inter-process communication, that represents a major issue for 
the parallel optimization of the standalone CFD and DEM parts, a generic coupling features an additional inter-physics 
inter-process communication channel. This can lead to massive inter-process communication of the coupling, that can deeply penalize
its performance.
 
In order to cope with the problem of memory consumption, in~\cite{KafuiParallel}, the authors proposed an optimized parallelization 
strategy for the DEM part, and an independent parallelization for the CFD part, that allowed reducing memory consumption 
while keeping an intensive inter-physics communication. 
The resulting coupling was observed to scale better than what previously seen in the literature
($\sim$35 for 64 processes) still performing 
significantly worse than the sole DEM part.
This proved how the inter-physics communication between the 
Eulerian and the Lagrangian part can induce major performance issues.

In ~\cite{EfficientOpenMPICFD-DEM}, the authors tried to cope with this problem by adopting a communication strategy based 
on non-distributed memory. The results proved 
how the reduced communication cost can indeed help the parallel performance of the coupling, yet its advantages are limited
to the usage of $\sim$30 processes, and therefore, not suitable for very-large-scale simulations. 

An  attempt to tackle large-scale simulations with an Eulerian-Lagrangian coupling was proposed in ~\cite{SediFoam}.
In~\cite{SediFoam}, the authors propose a coupling between the LAMMPS code, originally developed for  molecular dynamics applications,
and the OpenFOAM{\textregistered} libraries. In that work, the decomposition strategies for CFD and molecular dynamics code were completely independent from 
one another, and all the required data for the inter-physics exchange was provided through a complex communication layer.
It was shown that the specific couple of codes can address large-scale problems and  maintain good scalability properties when operated on one hundred of processes.
On the other hand, when a larger number of processes is used, the inter-physics communication becomes very important, taking up to 
30\% of the computation time and creating a major bottleneck for the parallel execution.

In~\cite{Pozzetti-co-located}, we proposed a way to overcome such a bottleneck. This strategy consists of imposing a 
co-location constraint between the partitions of the CFD and DEM domains so to perform the inter-physics exchange locally.
This strategy was proven to solve the bottleneck presented in~\cite{SediFoam}. Nevertheless, it reduces the flexibility 
in the partitioning of the domain, as configurations violating the co-location constraint are no longer possible.
In~\cite{Pozzetti-co-located}, we pointed out how this limitation would be more severe in case of non-uniformly distributed 
loads.

In this contribution, we propose a parallelization strategy that takes advantage of a parallel implementation of the grid-interpolation
 executed in the dual-grid multiscale approach, to overcome the shortcomings of the co-located
partitioning strategy while keeping its advantages.

The paper is structured as follows: First, the main structure of a dual-grid multiscale CFD-DEM coupling are proposed with 
focus on the communication required between the DEM element and the CFD grids.
Second, the partitioning strategy for the parallel execution of the coupling is proposed and the different communication
layers are identified.
Third, the grid-based parallel  interpolator is proposed in its simpler version.
The CFD and DEM equations resolved in section~\ref{Tests} are then briefly recalled.
Finally, three benchmark cases are proposed to assess the consistency and performance of 
the parallelization strategy.

We claim that the proposed strategy solves the parallelization issue underlined in~\cite{PozzettiIJMF}, and at the same time
overcomes the lack of flexibility that was affecting~\cite{Pozzetti-co-located}.

\section{Method}
\subsection{Dual-Grid Multiscale CFD-DEM coupling}
\begin{figure}[ht!]
\centering
 \begin{tikzpicture}[node distance=2cm]
 \node (coarsemesh) [region] {Coarse Grid};
 \node (descrCoarse) [process, right of=coarsemesh, xshift=1.65cm] 
{-Mapping Lagrangian fields to Eulerian.\\ -Solving fluid-particles 
interactions};
 \node (mapping) [ultra thick, draw=red, ellipse, minimum width=10pt,
    align=center, right of=coarsemesh, xshift=1.65cm, yshift=-3cm] 
{Grid-to-Grid\\ Interpolation};
 \node (fineMesh) [region , below of=coarsemesh, yshift=-3.9cm] {Fine Grid};
 \node (descrFine) [process, right of=fineMesh, xshift=1.5cm] {- Solving fluid  flow
 equations};
 \draw[-latex] (coarsemesh) to[bend right=10] node[above,rotate=90] {Particle 
related fields} (fineMesh);
 \draw[-latex] (fineMesh) to[bend right=10] node[below,rotate=90] {Fluid 
Solution} (coarsemesh);
\end{tikzpicture}
 \caption{\label{MULTISCALEWORKFLOW} Diagram of the solution procedure of the 
two different length scales for the 
simulation. The two boxes represent the different models, while the 
arrows show schematically the communication between the 
scales.
A coarse grid (top) is used to map the Lagrangian field of the DEM into an 
Eulerian reference and to solve the fluid-particle interaction.
Particle-related fields are mapped to the supporting domain (bottom) where 
a finer grid is used to solve the fluid equations.}
\end{figure}
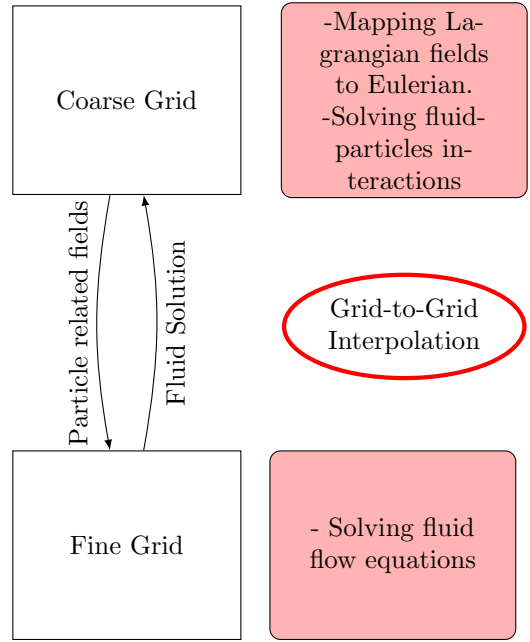

The concept of dual-grid multiscale approach for CFD-DEM couplings was introduced in~\cite{PozzettiIJMF}. 
In a dual-grid multiscale CFD-DEM coupling, two length-scales are identified within the particle-laden flow: a bulk scale at which the 
fluid-particle interaction is resolved, and a fluid fine-scale at which the equations governing the fluid 
phase are discretized.
As shown in figure~\ref{MULTISCALEWORKFLOW}, a coarse and uniform grid is adopted to resolve the fluid-particle interactions at the bulk scale, and a fine grid 
is used to solve the fluid flow equations at the fine-scale. An interpolation strategy ensures the communication between the two grids
i.e. the two lengthscales.

The dual-grid multiscale approach was originally introduced with reference to the coupling between DEM and VOF
as the fine resolution required by the VOF scheme was found to induce a marked separation between the two scales.
Nevertheless, this coupling strategy can be applied to a generic CFD-DEM coupling being most beneficial when 
such a scale separation is present.

From a software point of view, the implementation of the dual-grid multiscale approach requires the presence of 
two computational grids and a routine to interpolate between them, as depicted in figure~\ref{MULTISCALEWORKFLOW}.
The first implementation of the dual-grid multiscale approach used the \textit{OpenFOAM-extend} $meshToMesh$ library to perform
this interpolation. In the present contribution, we adopt a modified version of the library as discussed in section~\ref{MeshToMeshNew}
to perform the interpolation between grids in parallel.

In~\cite{PozzettiIJMF} it was pointed out how the coarse and fine grids should be independent from one another in order to 
allow maximum flexibility of the approach. For instance, while an uniform grid is optimal when performing a
Eulerian-Lagrangian coupling, a nonuniform, locally refined, computational grid provides several advantages for the solution
of complex flows. For this reason, the two grids will not, in general be nested.
When operated in parallel, the interpolation between the two grids will therefore require inter-process communication, as 
the coarse and fine domain partitions will not be, in general, perfectly aligned.
This might represent a problem as the introduced layer of intra-physics communication 
(i.e. the interpolation between bulk and fluid fine scale) could potentially negatively affect the parallel performance of the software.
As proposed in section~\ref{DomDeco}, the main idea of the current work is to take advantage of this parallel 
communication so to avoid inter-physics inter-process communication completely.

\subsection{Domain Decomposition and Multiscale Parallel communication}\label{DomDeco}
\begin{figure}[ht!]
\centering
\includegraphics[width=0.9\textwidth]{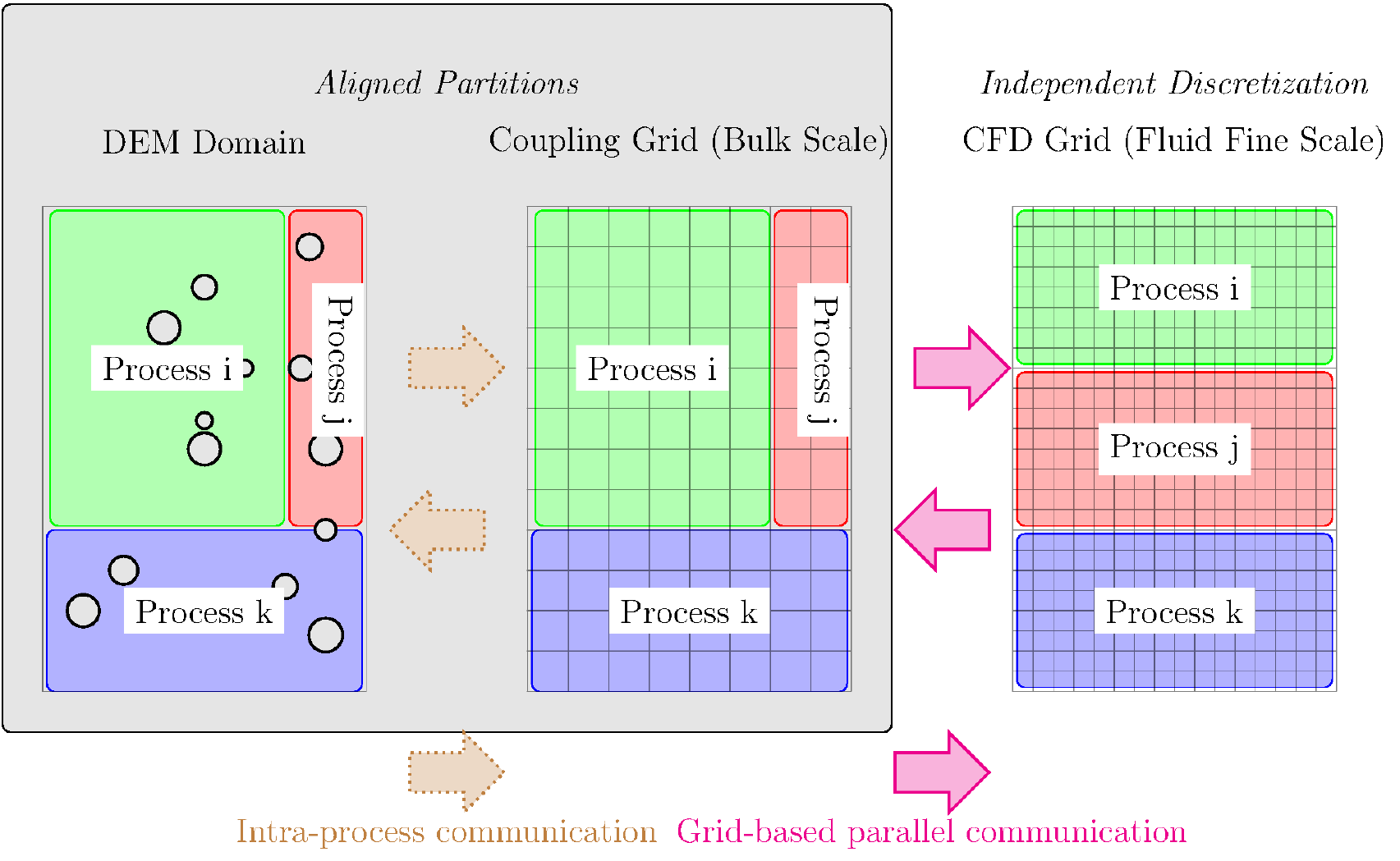}
\caption{\label{CFD-DEMPart}CFD-DEM domain partitioning and communications.}

\end{figure}
In~\cite{Pozzetti-co-located} was shown how using co-located partitions  for the CFD and the DEM domains provides significant advantages in the 
parallel execution of the coupling in terms of both memory consumption and reduction of inter-process communication.
In particular, it allows overcoming the inter-physics communication bottleneck that was identified in the 
literature~\cite{SediFoam}.

This bottleneck was caused by the coupling operation that retrieves information from the CFD-grid to the particle position and 
averages the particle-related quantities to define equivalent Eulerian fields. 
Since the Lagrangian entities continuously change their position, the connectivity between the particles and the CFD grid needs to be 
updated at every coupling step. When CFD and DEM information is stored on different processes, this implies the creation of a new 
data structure that must be sent via MPI messages. When this is done for computationally expensive cases and a high number of 
processes, the coupling procedure can become the most expensive part of the simulation.

As shown in~\cite{Pozzetti-co-located}, this problem can be  solved by partitioning CFD and DEM domain in such a way that all
the information required for the inter-physics exchanges will always be local. This 
allows keeping the communication time negligible also when operating over hundreds of processes.
Nevertheless, the approach proposed in~\cite{Pozzetti-co-located} suffers from a lack of flexibility in the domain partitioning, that can represent 
a main issue for cases featuring a non-uniform distribution of the CFD and DEM parts.

As shown in figure~\ref{CFD-DEMPart}, we propose to adopt a partitioning strategy based on perfectly aligned co-located partitions between the DEM domain and the coarse
grid. In this way, we ensure that the exchange of information between Lagrangian and Eulerian fields will be performed within 
the same process using standard library calls as in~\cite{Pozzetti-co-located}.

On the other hand, we rely on a parallel communication for the interpolation between the coarse and the fine grid. In this way, we allow 
a higher flexibility in the partitioning of the fine CFD grid, to which the computational load of the CFD is associated.
This grid does not need to be aligned with the DEM one and will, therefore, allow 
more complex partitioning strategies.
This introduces a higher communication cost if compared to the perfect alignment proposed in~\cite{Pozzetti-co-located}, 
but differently from what was happening in~\cite{SediFoam}, the grid-based communication is much
simpler than the Eulerian-Lagrangian one, as it relies on a static data-structure that, for non-deforming grids, can be calculated once per simulation.

\subsection{Grid-based parallel communication}\label{MeshToMeshNew}

\begin{figure}
\begin{minipage}[b]{.3\textwidth}
\includegraphics[width=0.9\textwidth]{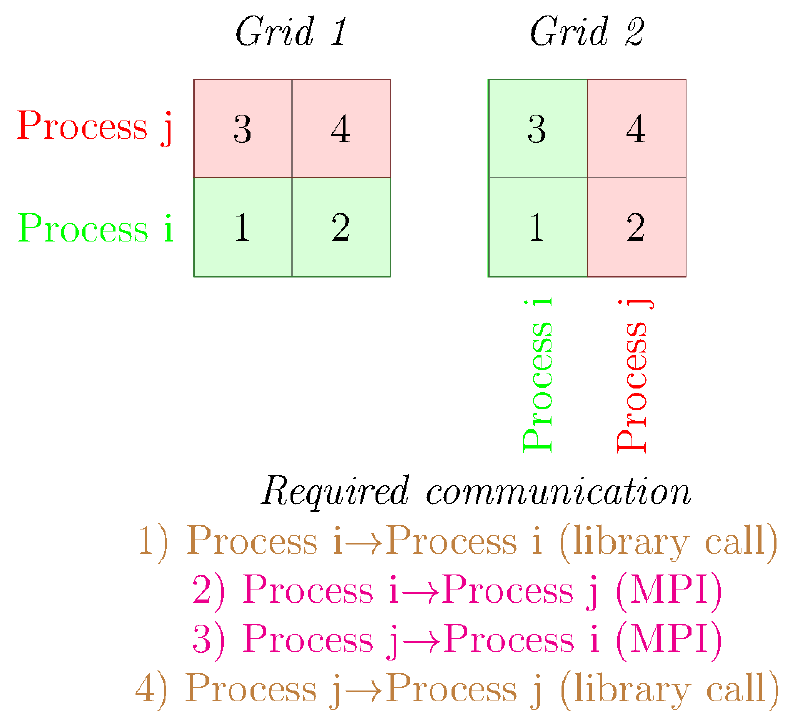}\subcaption{\label{grid_dummy}}
\end{minipage}
\begin{minipage}[b]{.65\textwidth}
\includegraphics[width=0.9\textwidth]{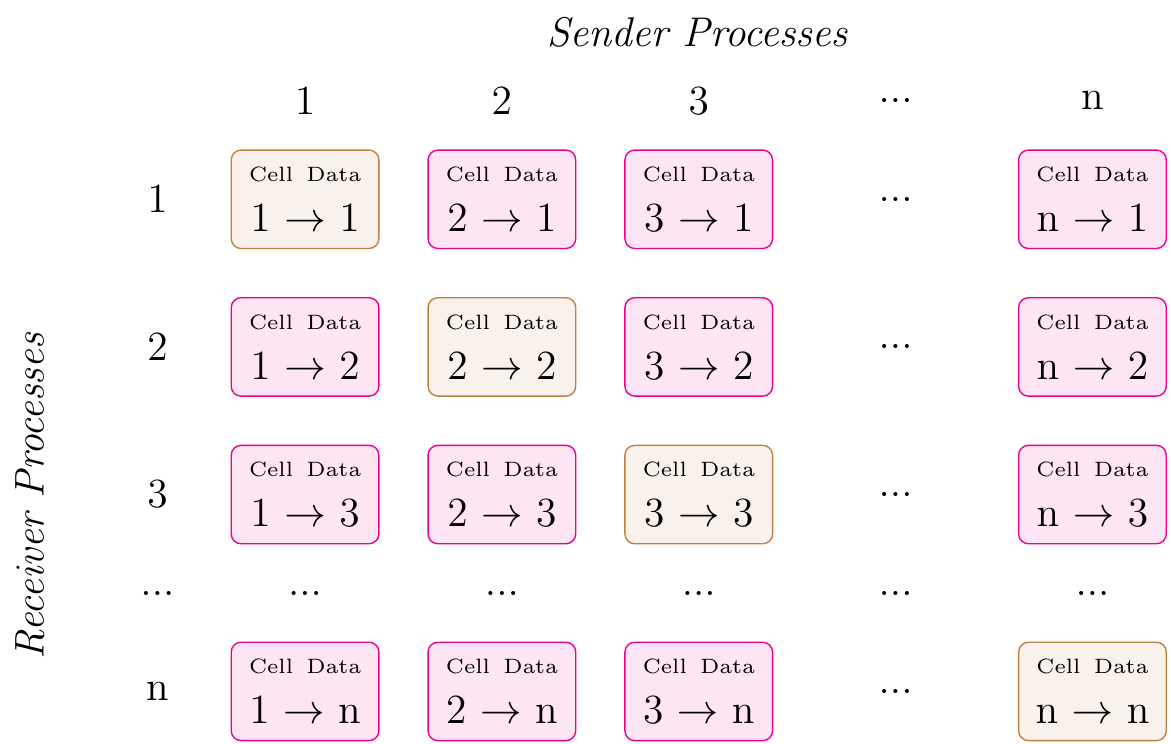}\subcaption{\label{matrix_general}}
\end{minipage}
\caption{\label{GridCommDummy}Grid-based parallel communication: two identical grids of 4 elements distributed on two different partitions~\ref{grid_dummy},
and data structure of a generic grid-based communication~\ref{matrix_general}.}
\end{figure}

\begin{figure}
\begin{minipage}[b]{.5\textwidth}
\includegraphics[width=0.9\textwidth]{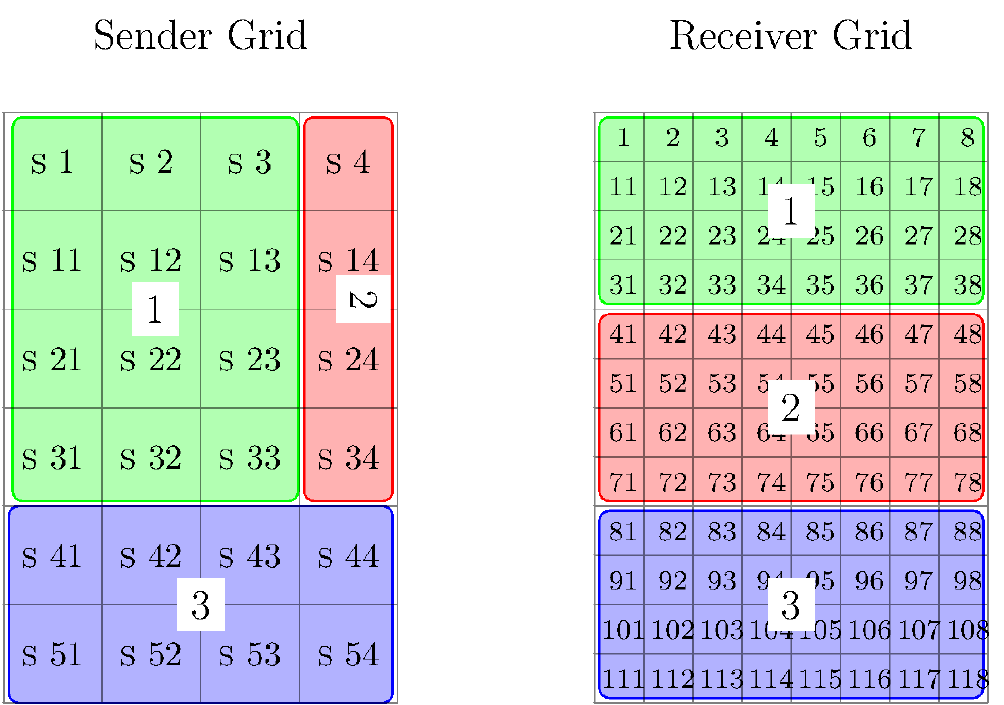}\subcaption{\label{grid_example}}
\end{minipage}
\begin{minipage}[b]{.5\textwidth}
\includegraphics[width=0.9\textwidth]{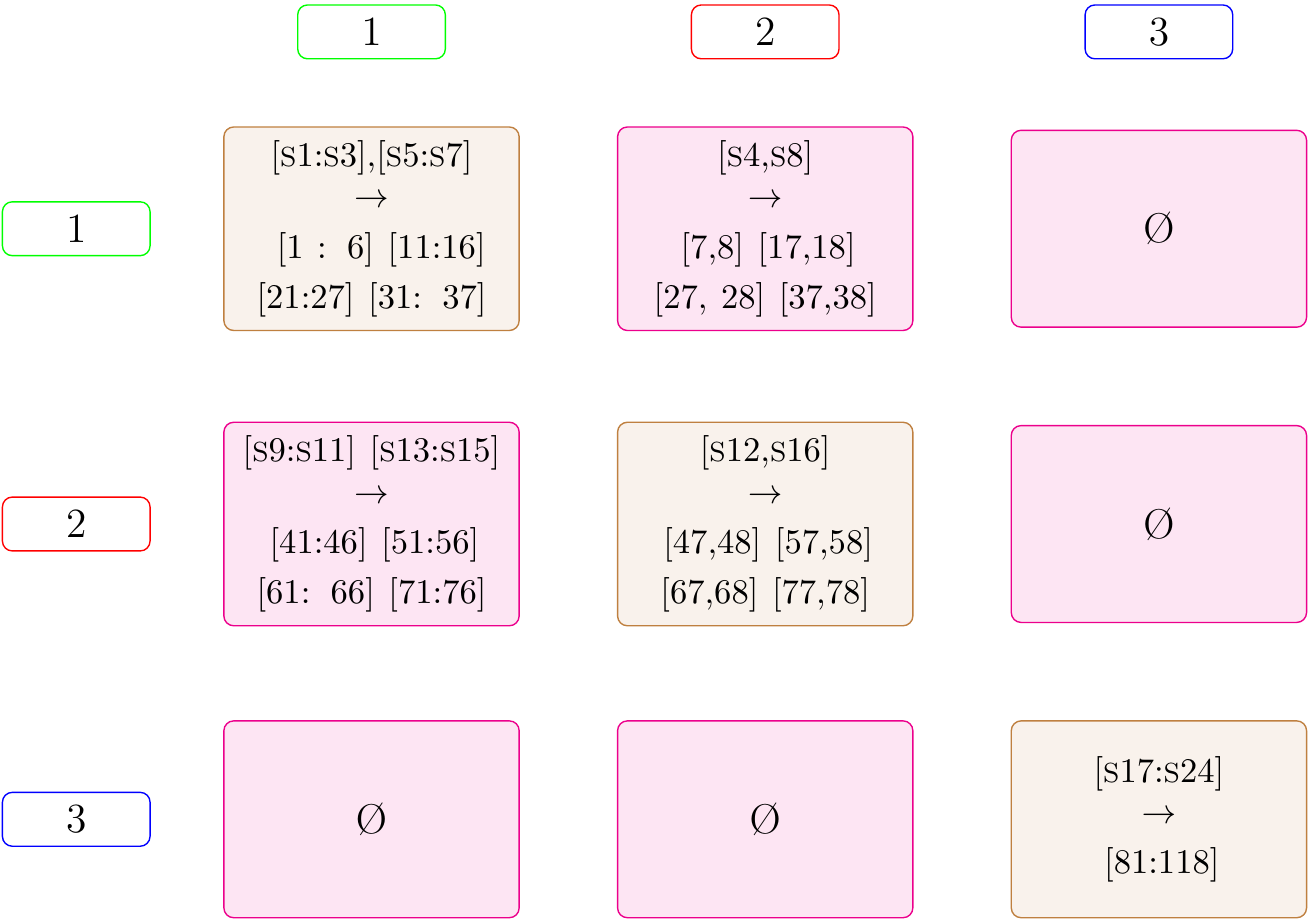}\subcaption{\label{matrix_example}}
\end{minipage}

 \caption{\label{Communication_example} Grid-based parallel communication: Example of a more complex grid partitioning~\ref{grid_example} and the relative communication matrix~\ref{matrix_example}.}
\end{figure}
During a parallel execution, the coarse and fine grids are partitioned into sub-domains that are distributed
in different processes. At runtime, every process stores a sub-domain of the coarse grid and a sub-domain of the fine grid.
For the sake of simplicity, we now refer to a one-way interpolation where one grid holds the updated information and one grid
should receive the interpolated values. We will refer to the first grid as the sender grid and to the latter as the receiver grid.
When performing the interpolation between grids, the information can be local (stored in the same process), non local (present in another
process) or a combination of the two.
We will here refer to a point-to-point communication assuming all the interpolations to retrieve the values to send, and to distribute 
the received values, to happen locally.

In figure~\ref{grid_dummy}, we propose a simple example of two identical grids of  4 elements distributed on two processes \textit{i} and \textit{j}.
In this case, the interpolation will consist of a simple map of the values from one cell of the sender grid
to its twin cell of the receiver. 
In order to map all the 4 values from the sender grid (\textit{Grid 1}) to the receiver grid (\textit{Grid 2}), four communications are needed.
For the communication 1 and 4, the information is local, therefore the operation can be done with simple library calls. 
On the other hand, for communication 2 and 3, the sender and receiver elements are located on different processes. Therefore, those communication
must be performed via MPI.

In figure~\ref{matrix_general}, we propose a representation of the data-structure required for the grid-communication in a more general case 
of two meshes distributed over n processes. An element (\textit{i,j}) of the matrix represents a dataset that needs to be communicated from process
\textit{j} to process \textit{i}. The diagonal terms represent the intra-process communication that are executed via library calls, while the 
non-diagonal terms require parallel communication. If the two meshes do not have relative movement over time (as is normally happening in CFD-DEM couplings),
the same data-structure will hold during all the simulation, and must be calculated only once. An example of this data structure for a 
more realistic grid-partitioning is presented in figure~\ref{Communication_example} were it can be observed how, in general, not all 
the elements \textit{i,j} will contain data that need to be communicated, and therefore the communication matrix will be sparse.

With the rise in the number of processes, the matrix generally becomes sparser and sparser and, therefore, not all the elements
will, in general, contain a dataset. Furthermore, every process will only have to write the kth column and only have to read from the 
kth row, so that a global knowledge of the data contained in this object is not needed. In particular, the diagonal term (\textit{i,i})
is never required by process \textit{j}.

In this work, we compare two strategies for performing the parallel communication to which we will refer to as \textit{gather-scatter} and 
\textit{distributed}.
In a \textit{gather-scatter} strategy, every process holds knowledge of the non-diagonal part of the matrix of figure~\ref{matrix_general}.
During the mapping operation, every process \textit{k} writes the kth columns, then all the information is gathered from the master node and 
 scattered back to all the processes. Afterwards, the process \textit{k} reads the kth row and stores the interpolated values on the local sub-domain
of the receiver grid. This strategy was chosen for its simplicity and to provide a reference to study the importance of the grid interpolation on the 
parallel performances of the coupling.
In a \textit{distributed} strategy, a process \textit{k} only holds knowledge of the k-th column and the k-th row of the matrix and 
directly communicates the element \textit{k,j}(if existing) and receive the element \textit{j,k}(if existing) from the process \textit{j}.
In this way, the dataset (\textit{k,j}) will be sent form process \textit{k} to process \textit{j} without passing from the master.
In section~\ref{10M10M} we compare the performance of the two methods to underline when a more complex parallel communication is needed.

\subsection{Equations solved in the DEM Domain}
\begin{table}[tbp]
\centering
\caption{List of variables}
\label{my-label}
\begin{tabular}{p{1cm}p{5cm}p{1cm}p{5cm}}
\arrayrulecolor{blue}
\cline{1-4}
Symbol &  Variable & Symbol &  Variable  \\
\cline{1-4}
\\
 $\mathbf{u}_f$ &  fluid velocity &  $\alpha $ & volume fraction of the liquid phase\\ 
 $\mu_f$ &  fluid viscosity & $\mathbf{u}_c$ &  compression velocity\\
 $\rho_f$ &  fluid density & $\Gamma$ & Surface Tension\\
 $\mathbf{F_{fpi}}$ &  volumetric source of momentum due to interaction with 
particles  &  $\mathbf{T_{\Gamma}}$  & surface 
tension force \\

 $\mathbf{F_{b}}$ &  body force &$\epsilon$   & porosity field:$\sum_i V_{pi} /V_{cell}$ with $i =1.... n_{pc}$\\
 $p$   & pressure &  $\chi$ & phase indicator\\

\cline{1-4}\\
 $\rho_p$ &  particle density &$\mathbf{F}_g$ &  gravitational force\\
 $d_p$ &  particle diameter &$\mathbf{M}_{coll}$ &  torque acting on a particle due to collision events \\ 
 $m$ &  particle mass &$\mathbf{M}_{ext}$ &  external torque acting on a particle\\
 $\mathbf{I}$ &  particle moment of inertia & $\mathbf{u}_p$ &  particle velocity\\
 $\mathbf{F}_c$ &  contact force acting on a particle &$A_p$ &  particle surface area\\ 
 $\mathbf{F}_{ij}$ &  contact force acting on the particle $i$ due to the collision with particle $j$ & $\bm{\phi}$ &  particle orientation\\
 $\mathbf{F}_{drag}$ &  force rising from the particle-fluid interaction &$\bm{\omega}$ & particle angular velocity \\ 

 $\bm{n}$ & particle-fluid interface normal vector &$\beta$ &  drag factor\\

\cline{1-4} 
\end{tabular}
\end{table}
In this work, we rely on the dynamic module of the XDEM Platform~\cite{PozzettiIJMF,BaniasadiMelting,bpadditivemanufacturing}, to evolve a set of discrete entities
moving in the presence of a multiphase flow.
In this section, we briefly recall the equations resolved by this module and the structure of domain-partitioning 
adopted within XDEM.
The positions and orientations of the particles are updated at every time-step according to
\begin{eqnarray}\label{DEM_EQUATIONS}
m_i \frac{{d}^2}{{dt}^2}\mathbf{x}_i & =& 
\mathbf{F}_{coll}+\mathbf{F}_{drag}+\mathbf
{F}_g , \\
\mathbf{I}_i \frac{{d}^2}{{dt}^2} \bm{\phi}_i & =& 
\mathbf{M}_{coll} + \mathbf{M}_{ext},
\end{eqnarray}
where $\mathbf{x}_i$ are the positions, $m_i$ the masses, and $ \bm{\phi}$ the orientations of the entities.
The term $\mathbf{F}_c$  indicates the force arising from collisions
\begin{eqnarray}\label{FORCES}
\mathbf{F}_{coll} = \sum_{i\neq j} \mathbf{F}_{ij}( \mathbf{x}_j, \mathbf{u}_j, 
\bm{\phi}_j, \bm{\omega}_j ),
\end{eqnarray}
{with  $\mathbf{u}_j$ the velocity of particle $j$, and $\bm{\omega}$ the angular 
velocity.
The term  $ \mathbf{M}_{coll} $ represents the torque acting on  
the particle due to collision{s}
\begin{eqnarray}\label{TORQUES}
\mathbf{M}_{coll} = \sum_{i\neq j} \mathbf{M}_{ij}(\mathbf{x}_j, \mathbf{u}_j, 
\bm{\phi}_j, \bm{\omega}_j ),
\end{eqnarray} 
with $\mathbf{M}_{ij} $ the torque acted from particle $j$ to particle $i$.
The term $\mathbf{F}_{drag}$ takes into account the force rising from the 
interaction with the fluid, 
and $\mathbf{F}_g$ corresponds to the gravitational force.
For what concerns $\mathbf{F}_{drag}$ a semi empirical model 
in the form :
\begin{eqnarray}\label{DragLaw}
\mathbf{F}_{drag}= \beta (\mathbf{u}_{f}-\mathbf{u}_{p}),\\
\beta=\beta(\mathbf{u}_{f}-\mathbf{u}_{p},\rho_f,\rho_p,d_p,A_p,\mu_f
, \epsilon),
\end{eqnarray}
is here chosen, with $\mathbf{u}_{f}, \, \mathbf{u}_{p}$ the fluid and particle 
velocity respectively,  $\rho_f,\,\rho_p$ the respective densities, $d_p,\,A_p$ 
the 
particle characteristic length and area, $\mu_f$ the fluid viscosity, and  
$\epsilon$ the porosity, defined as the ratio between the volume occupied by the 
fluid and the total volume of the CFD cell.
For the sake of generality, we took $\beta$ as described in~\cite{New1}.

As described in~\cite{Pozzetti-co-located}, in the parallel execution of XDEM, the simulation domain is geometrically 
decomposed in regularly fixed-size cells that are used to distribute the workload between the processes.  
In this way, every process will be assigned a set of cells that will define its sub-domain.
Every process only performs the calculation and holds knowledge
of the particles that belong to its sub-domain.

Different load-partitioning algorithms are available within the platform XDEM~\cite{Pozzetti-co-located}, 
among which a dedicated partitioner that is able to force a co-location constraint between XDEM and OpenFoam partitions as
proposed in~\cite{Pozzetti-co-located}.

\subsection{Equations solved in the CFD Domain}

For the solution of the fluid flow equations, we rely on the OpenFOAM-extend libraries.
We adopt the solver presented in~\cite{PozzettiIJMF,PozzettiICNAAM2016} for tackling 
the general case of an unsteady incompressible multiphase flow through porous
media with forcing terms arising from the particle phase.
We here briefly recall the sets of equations governing such a system.

For an incompressible flow through porous media, the Navier-Stokes equations take the form 
\begin{eqnarray}\label{UFP}
\frac{\partial \epsilon \rho_f\mathbf{u}_f}{\partial t}+ \nabla \cdot (\epsilon 
\rho_f\mathbf{u}_f\mathbf{u}_f)= -\epsilon \nabla p+
\nabla \cdot \left(\epsilon \mu_f\left( \nabla \mathbf{u}_f+ \nabla^T 
\mathbf{u}_f
\right) \right)+\mathbf{T_{\Gamma}}+\mathbf{F_b} +\mathbf{F_{fpi}},\\
\nabla \cdot \epsilon\mathbf{u}_f=-\frac{\partial \epsilon}{\partial \tau}.
\end{eqnarray}
with $\mathbf{u}_f$ the fluid velocity, $p$ the fluid pressure,  $\mathbf{F_b} $ a generic body force, $ \mathbf{F_{fpi}}$
the fluid-particle interaction force, that is the counterpart of 
$\mathbf{F}_{drag}$, which is here treated with the semi-implicit algorithm proposed in~\cite{ROBUSTCFDDEM}.
Local density and viscosity are dependent on the fluid phase and can be written 
in the form
\begin{eqnarray}
\rho_f(\mathbf{x})=\rho_1\alpha(\mathbf{x})+\rho_2 (1-\alpha(\mathbf{x})),\\
\mu_f(\mathbf{x})=\mu_1\alpha(\mathbf{x})+\mu_2 (1-\alpha(\mathbf{x})),
\end{eqnarray}
with $\alpha$ the {volume fraction} defined by
\begin{eqnarray}
 \alpha &=& \frac{1}{V}\int_V \chi(\mathbf{x}) d\mathbf{x},\\
 \chi & = & \left\{ \begin{array}{cc}
       1 & \mbox{if first fluid, }\\
       0 & \mbox{if second fluid.}
       \end{array} \right.
\end{eqnarray}
{The volume fraction} is considered a scalar transported by the fluid flow for 
which 
\begin{eqnarray}\label{phase_advection}
 \frac{\partial \epsilon \alpha}{\partial t}+ \nabla \cdot (\epsilon \alpha \mathbf{u}_f)+ \nabla 
\cdot (\epsilon \alpha (1-\alpha)  \mathbf{u}_c)=0,
\end{eqnarray} 
must hold, with $\mathbf{u}_c$ the relative velocity between the two-phases 
referred {to} as 
compression velocity. {The third} term is introduced in order to avoid an excessive 
numerical dissipation.

This set of equations is solved with the OpenFOAM libraries~\cite{OpenFOAM}, 
which parallelization is based on domain decomposition.
The CFD domain is split into sub-domains assigned to each process available at 
run time, over each of them a separate copy of the code is run.
The exchange of information between processes is performed at boundaries through 
a dedicated patch class as described in~\cite{OpenFOAM}.
According to~\cite{Pozzetti-co-located}, the partitioning of the OpenFOAM domain 
is performed through an unique XDEM-OF partitioner that aims to enforce the co-location
constraint. In particular, as proposed in section\ref{DomDeco}, the coarse grid is partitioned 
enforcing a perfect alignment with the XDEM sub-domains, while the fine grid is partitioned independently.

\section{Test Cases}\label{Tests}

We here propose three benchmarks to assess the validity and the scalability of our approach.
The results refer to the coupling between OpenFoam-extend 3.2 and XDEM as in~\cite{Pozzetti-co-located}.

The first benchmark, \textit{One particle traveling across a processes boundary} is presented in section~\ref{OneParticle}.
It tests the equivalence of the results between the sequential run, a parallel run as proposed in~\cite{Pozzetti-co-located}, and a 
parallel run with the current approach featuring an arbitrary discretization of the CFD grid. 
This is done to assess the validity of our approach and implementation 
by checking the continuity of the results when the particle travels across a process boundary
even without partitions alignment.

The second benchmark, \textit{Three-phase dam-break} proposed in section~\ref{Dam-Break}, 
re-proposes a famous benchmark featuring complex multiphase interactions and a non-uniformly distributed
computational load.
This is done to assess how the present parallelization approach preserves the specific
properties of the dual-grid multiscale DEM-VOF coupling in terms of accuracy and grid convergence.
Also, it shows the benefit of the current strategy in case of non-uniform load distribution in the domain.

The third benchmark, \textit{Ten Million Particles in ten Million cells} described in section~\ref{10M10M},
studies the parallel performance of a coupled solution in case of a heavy coupled case. 
This is done to show how our solution can handle highly costly simulations 
and at the same time allows resolving a main issue underlined in the literature linked to the inter-physics communication. 
Furthermore, the test-case aims to underline the importance of the parallel implementation 
of grid-based communication for large-scale computation.

The experiments were carried out using the $Iris$ cluster of the University of Luxembourg~\cite{VBCG_HPCS14}
which provides 168 computing nodes for a total of 4704 cores.
The nodes used in this study  feature a total a 128 GB of memory 
and have two Intel Xeon E5-2680 v4 processors running at 2.4 GHz, that is to say a total of 28 cores per node.
The nodes are connected through a fast low-latency EDR InfiniBand (100Gb/s) network organized in a fat tree. 
We used OpenFOAM-Extend 3.2 and XDEM version b535f789736, both compiled with Intel Compiler 2016.1.150
and parallel executions were performed using Intel MPI 5.1.2.150 over the InfiniBand network.

To ensure the stability of the measurement, the nodes were reserved for an exclusive access. 
Additionally, each performance value reported in this section is the average of at least hundred of measurements.
The standard deviation showed no significant variation in the results.

\subsection{One particle traveling across process boundaries}\label{OneParticle}
\begin{figure}[ht]
\centering
 \includegraphics[width=0.7\textwidth]{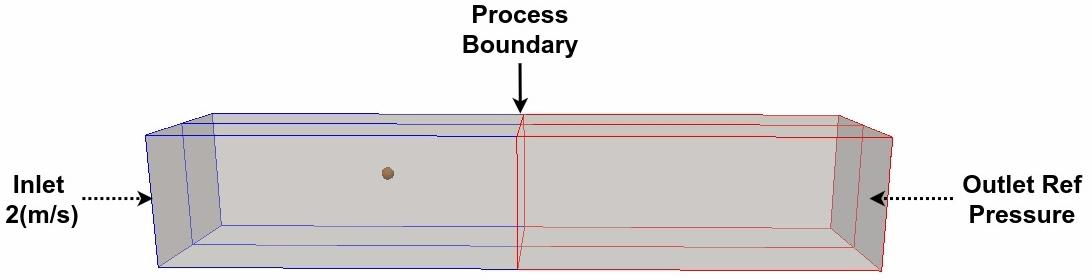}
 \caption{\label{setup} One particle traveling across processes boundaries: 
Setup.}
\end{figure}

\begin{figure}
\includegraphics[width=0.48\textwidth]{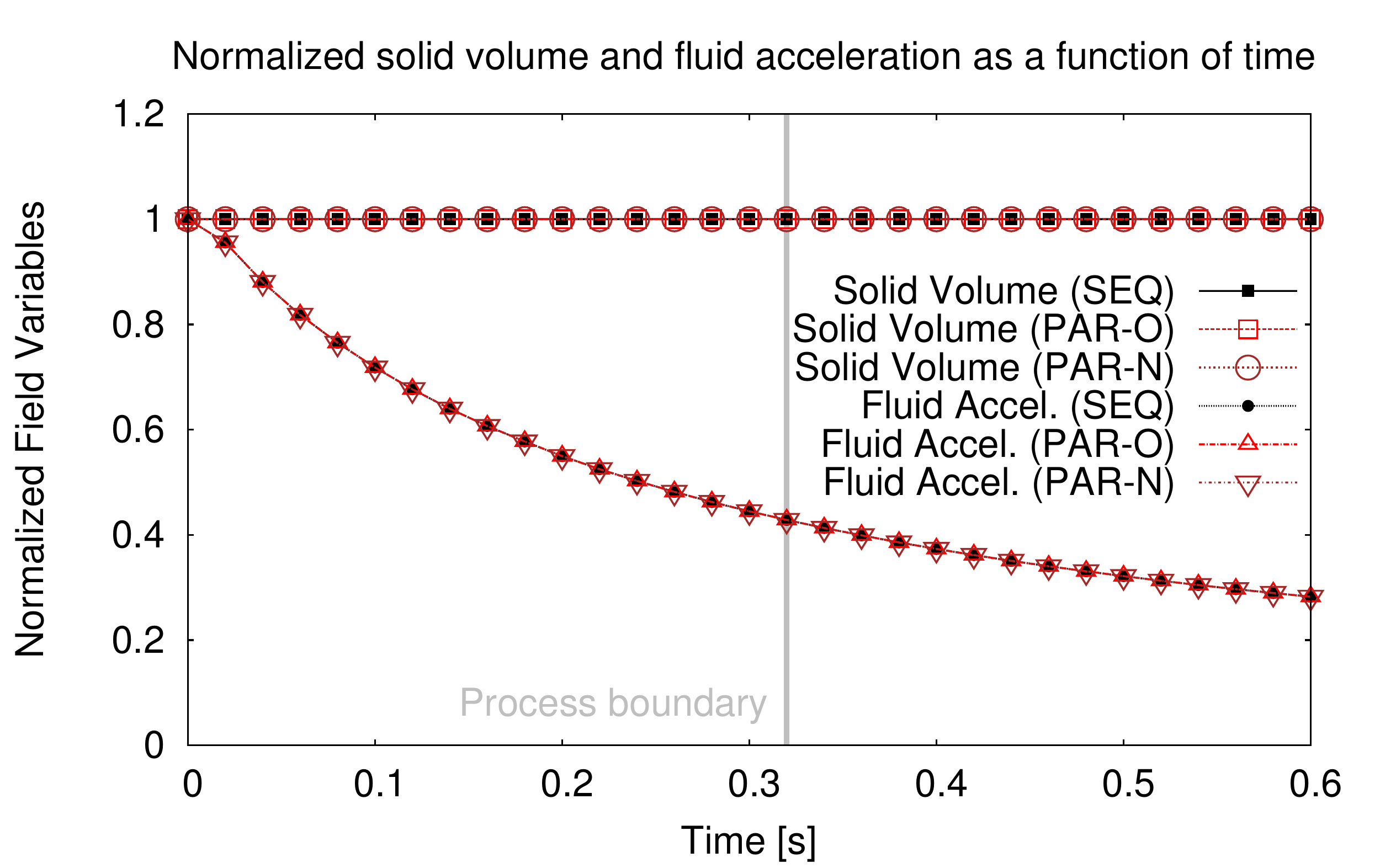}
\includegraphics[width=0.48\textwidth]{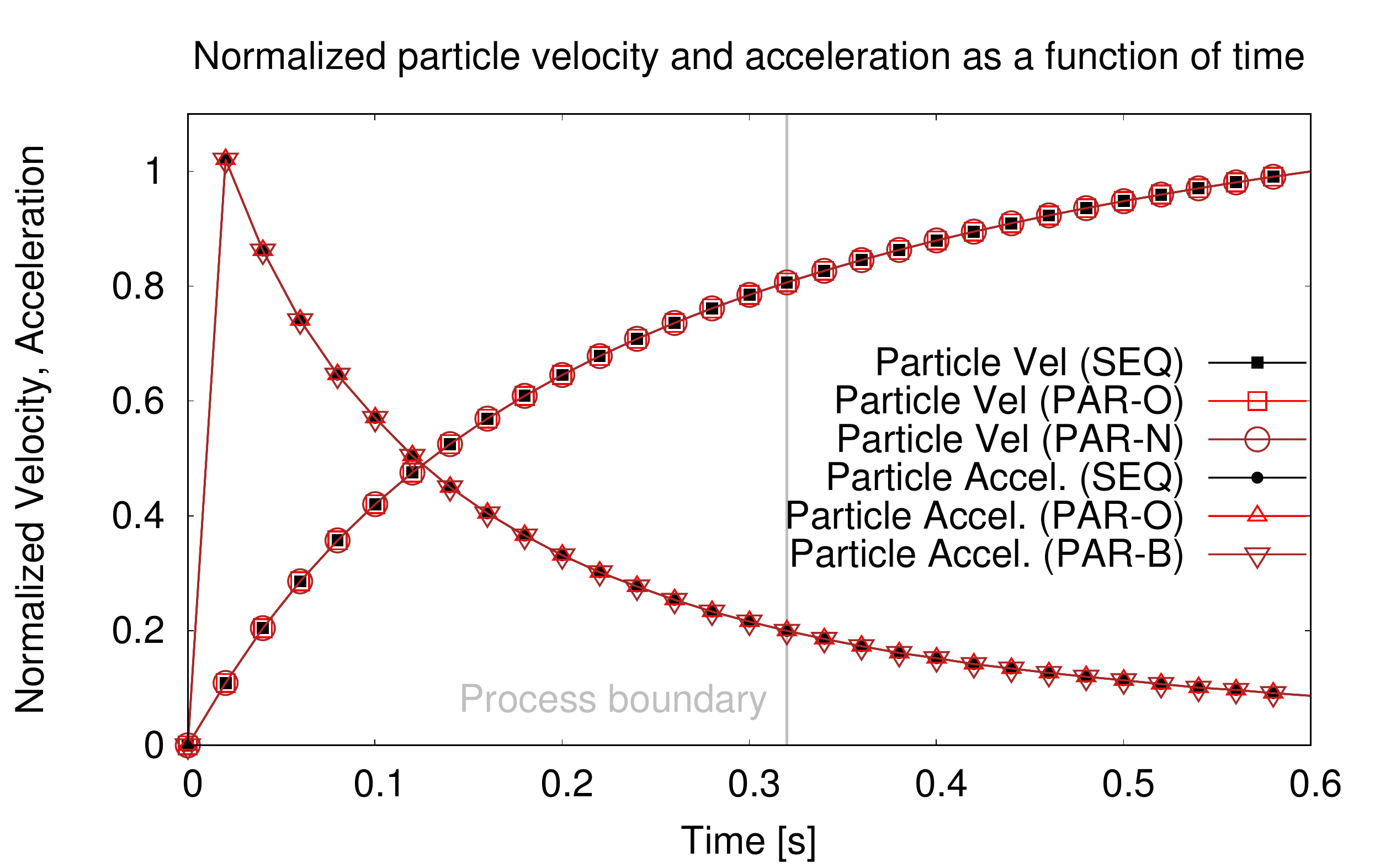}
\caption{\label{ParticleAcross} One-Particle traveling across processes 
boundaries:
 comparison between the results obtained with sequential execution (SEQ), parallel execution with overlapping partitions (PAR-O), and 
 parallel execution with non-overlapping partitions (PAR-N). Results for CFD 
(left) and DEM (right) 
variables.
 Continuity across processes boundary can be observed.}
\end{figure}

A first test is proposed, featuring a particle initially at rest traveling 
across a boundary between processes.
This is done in order to show the ability of the proposed approach to deal with arbitrary domain partitions
in which the information related to CFD and DEM parts can be stored in different processes.
The particle is initially at rest within the domain assigned to process $0$. The CFD fine-grid partition overlapping 
with the DEM partition assigned to process $0$ is here assigned to process $1$ so to mimic a situation of complete 
non-overlap of the DEM and CFD local domains.
The particle is 
accelerated by the fluid according to the law of equation \ref{DragLaw}. 
The resulting drag force pushes the particle across the boundary with process 
$1$ causing it to be transferred from the sub-domain $0$ to the sub-domain $1$ of the DEM part.
This represents the worse scenario for a coupling, in which none of the information required 
is local.

As shown in figure~\ref{setup}, the boundary conditions imposed on the fluid domain are an uniform Dirichlet at the inlet,
no-slip at the wall, and reference pressure at the outlet. 
The CFD domain is a square channel of dimensions of $1m$, $0.2m$ and $0.2m$, and is discretized with $240$ identical cubic cells. 

In figure \ref{ParticleAcross}, the normalized particle velocity and acceleration, and the normalized 
solid volume and $L_1$ norm of the fluid acceleration are proposed as a function of time.
All the quantities are normalized by dividing them for their maximum value, so that they can be displayed on the same plot.
The particle crosses the process boundary at $0.32 s$ leaving the DEM domain assigned to process $0$ and entering into the domain 
assigned to process $1$, and subsequently leaving the CFD domain assigned to process $1$ and entering into the domain assigned to process $0$.

It can be observed how the particle velocity and acceleration are 
continuous across the processes boundary.
This shows how the information on the fluid velocity at the particle position is correctly exchanged between
CFD and DEM code in the whole domain, including the regions between boundaries.
Similarly,  the porosity and acceleration fields projected by the particle into the Eulerian grid remain constant along time without suffering
discontinuities when the particle switches between processes.

In particular, the results are identical to the one obtained through a sequential execution
and a parallel execution with co-located partitions~\cite{Pozzetti-co-located}.
This shows how the proposed method allows overcoming the limitation induced by 
the standard co-located partitions strategy extending its applicability to more complex scenarios
where identical partitions for the two domains are not possible.

\subsection{Three-phase dam-break}\label{Dam-Break}
\begin{figure}[ht]
\centering
 \includegraphics[width=0.6\textwidth]{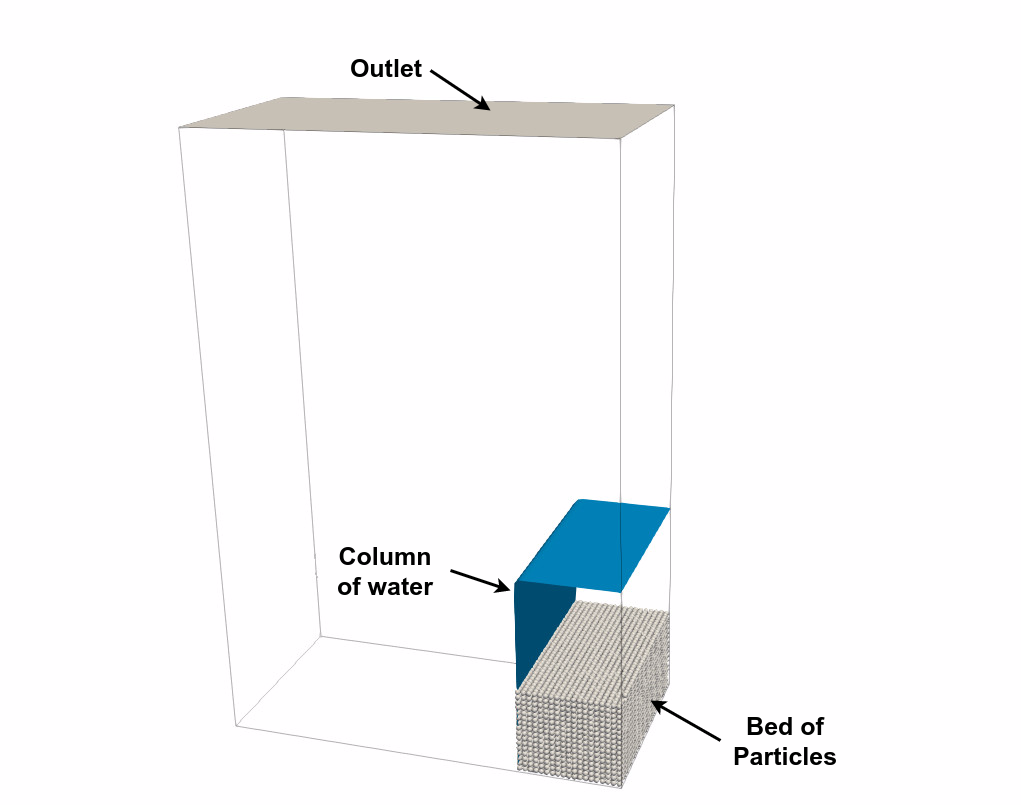}
 \caption{\label{Dam_Break_Setup} Three-phase dam-break. Simulation domain setup.}
\end{figure}
\begin{figure}
\begin{minipage}[b]{.5\textwidth}
\includegraphics[width=0.8\textwidth]{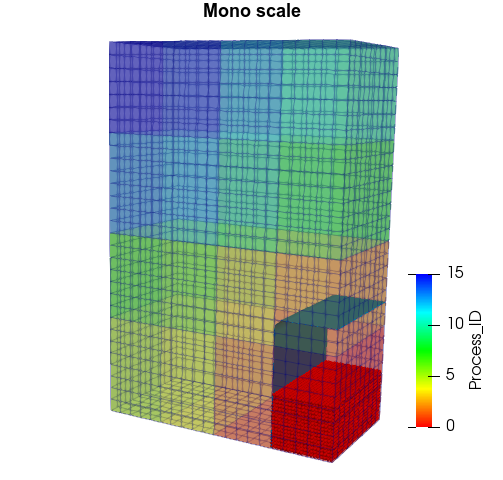}\subcaption{\label{MonoFull} Domain partitioning with a mono-scale approach.}
\end{minipage}
\begin{minipage}[b]{.5\textwidth}
\includegraphics[width=0.8\textwidth]{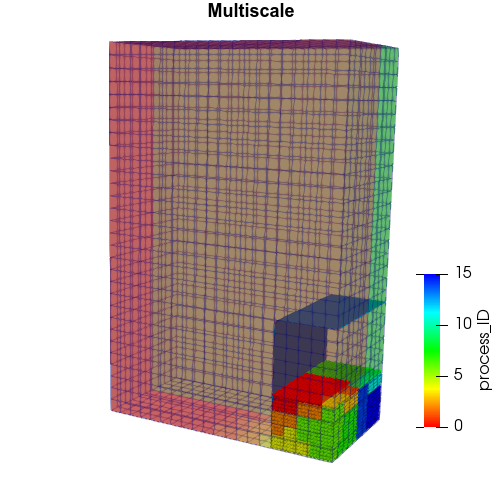}\subcaption{\label{MultiFull}Domain partitioning with a multiscale approach.}
\end{minipage}

\begin{minipage}[b]{.5\textwidth}
\includegraphics[width=0.7\textwidth]{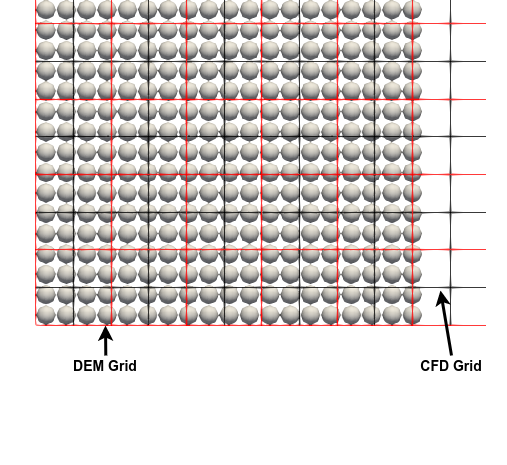}\subcaption{\label{MonoGrid} Grid discretization of the XDEM domain (red) and of the 
CFD domain (black) for the  mono-scale DEM-VOF coupling.}
\end{minipage}
\begin{minipage}[b]{.5\textwidth}
\includegraphics[width=0.7\textwidth]{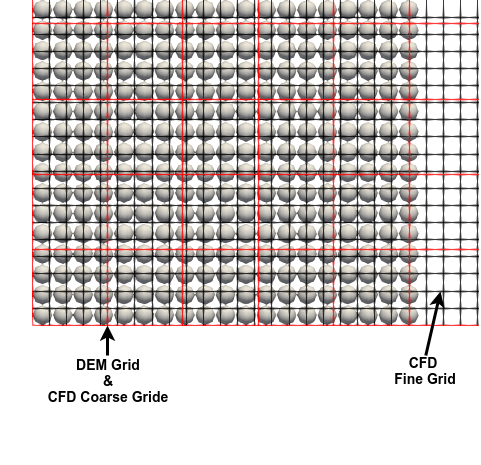}\subcaption{\label{MultiGrid}Grid discretization of the XDEM domain and CFD bulk scale (red) and of the 
CFD fine scale (black) for the  multiscale DEM-VOF coupling.}
\end{minipage}

\begin{minipage}[b]{.5\textwidth}
\includegraphics[width=0.8\textwidth]{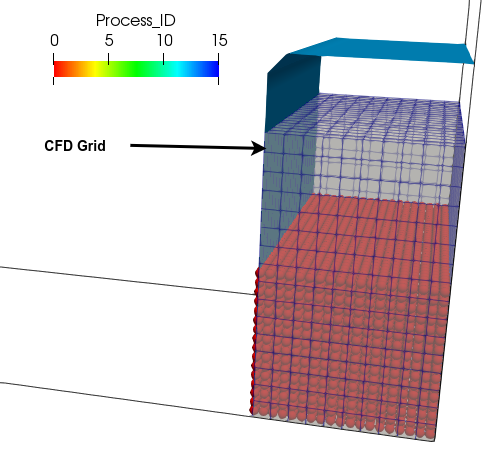}\subcaption{\label{MonoZoom}CFD and DEM partitions assigned to process 1  in a mono-scale coupling.}
\end{minipage}
\begin{minipage}[b]{.5\textwidth}
\includegraphics[width=0.8\textwidth]{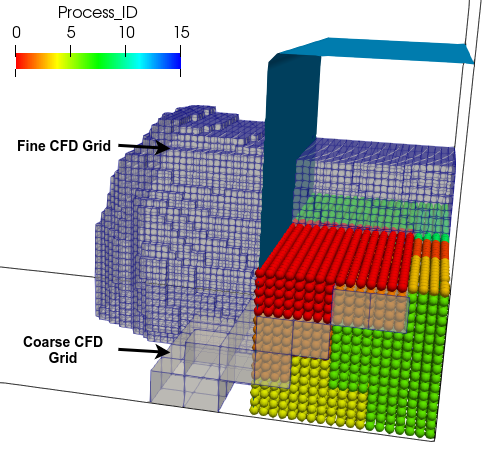}\subcaption{\label{MultiZoom}CFD and DEM partitions assigned to process 2 in a multiscale coupling.}
\end{minipage}
\caption{\label{Dam_Break_Partitioning} Three-phase dam-break. Simulation setup. Different possible partitioning of the domain with mono-scale  and 
 multiscale approach.}
\end{figure}

\begin{figure}[ht]
\centering
 \includegraphics[width=0.6\textwidth]{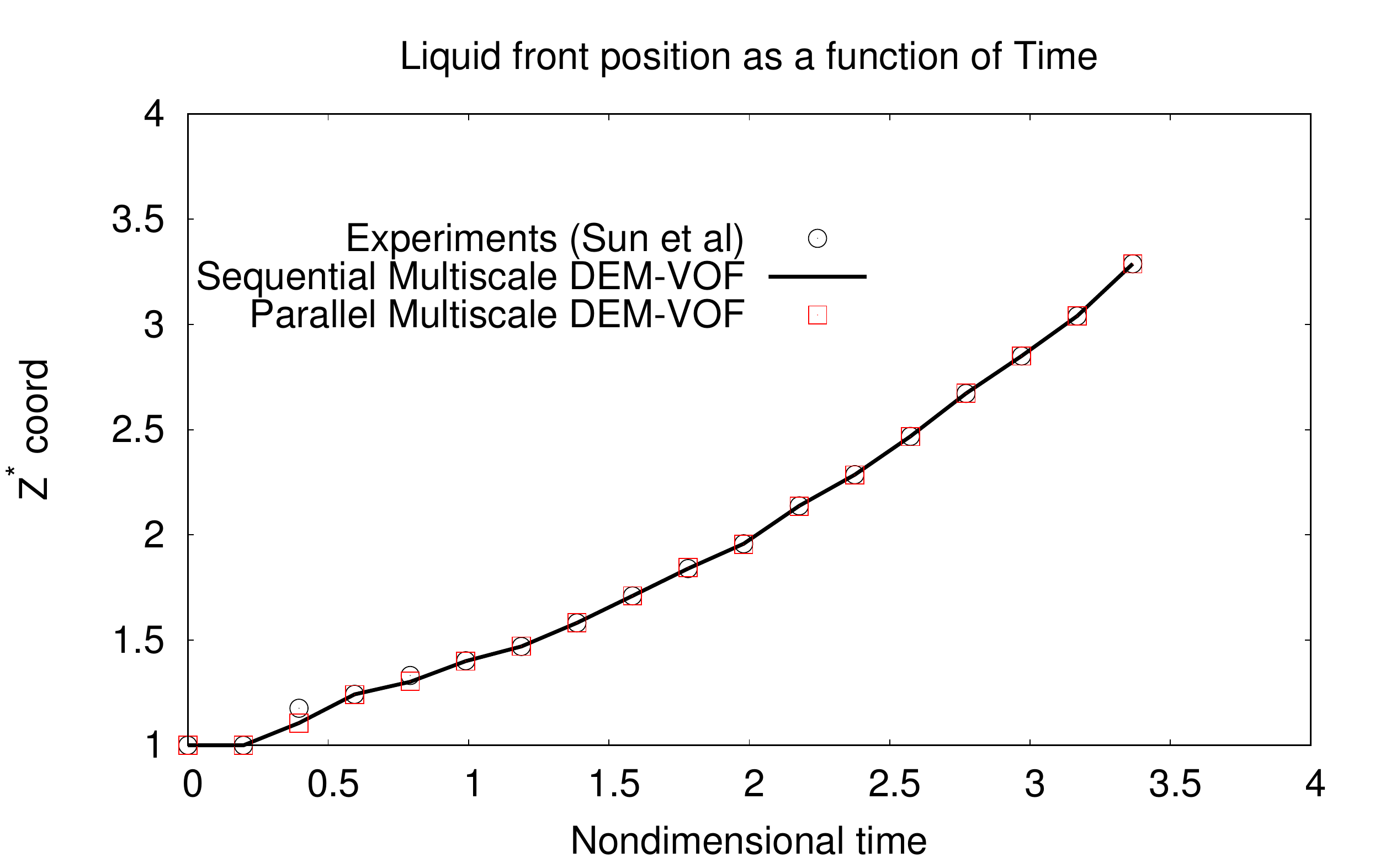}
 \caption{\label{Fronts} Three-phase dam-break. Liquid front position as a function of time. Comparison between Experimental data,
 sequential multiscale simulation and parallel multiscale simulation. Identical results can be observed betweer parallel and 
 sequential run, and good agreement is shown with the experimental results.}
\end{figure}
\begin{figure}[ht]
\centering
 \includegraphics[width=0.6\textwidth]{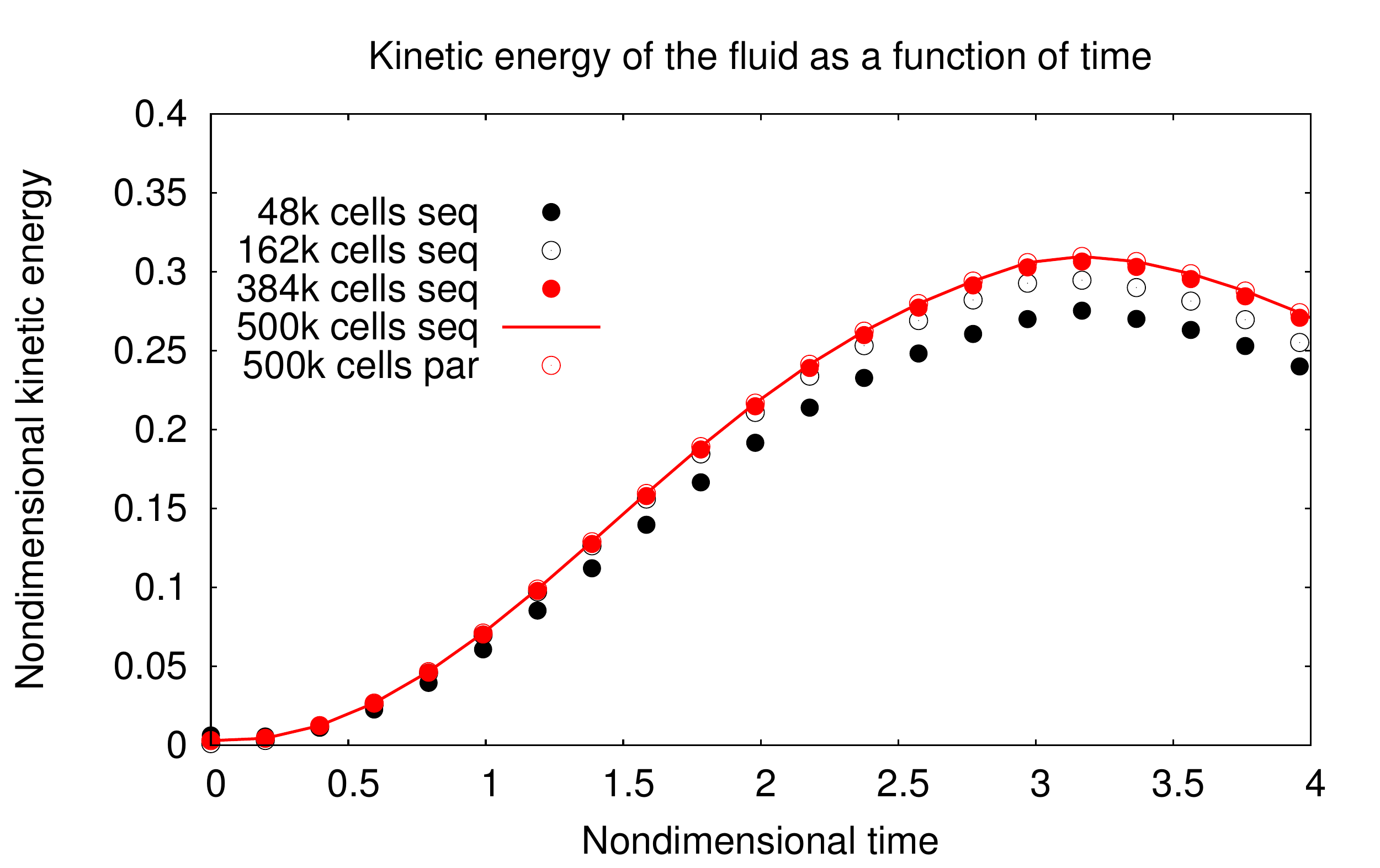}
 \caption{\label{Kin_en} Three-phase dam-break. Non-dimensional kinetic energy as a function of non-dimensional time. Comparison
 between sequential run with different fine grides resolution and parallel run with 500k discretization.}
\end{figure}

\begin{figure}[ht]
\centering
 \includegraphics[width=0.6\textwidth]{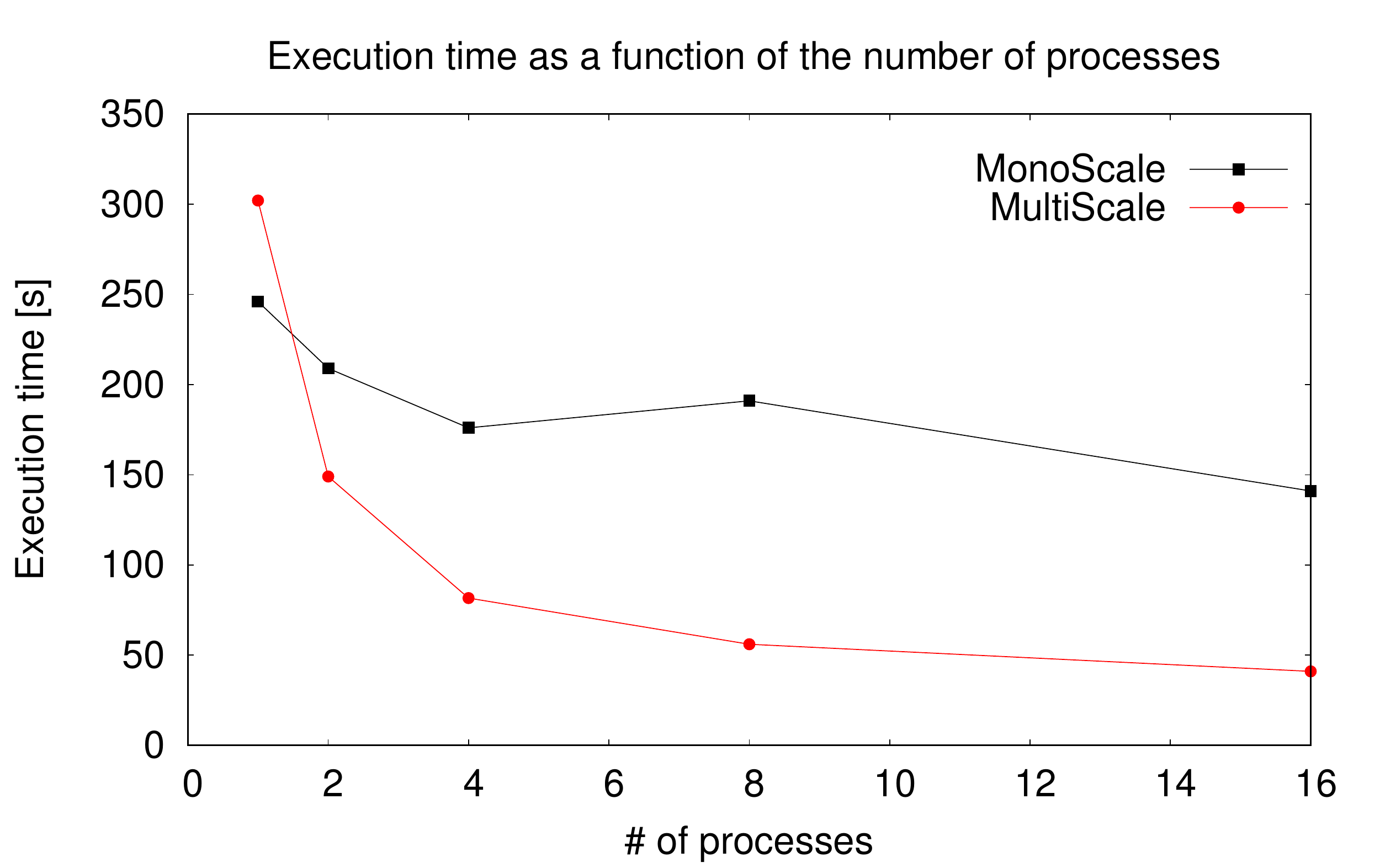}
 \caption{\label{DamBreakTime} Three-phase dam-break. Execution time as a function of the number of processes.
 better parallel performance is observed for the multiscale approach.}
\end{figure}

\begin{figure}[ht]
\centering
 \includegraphics[width=0.6\textwidth]{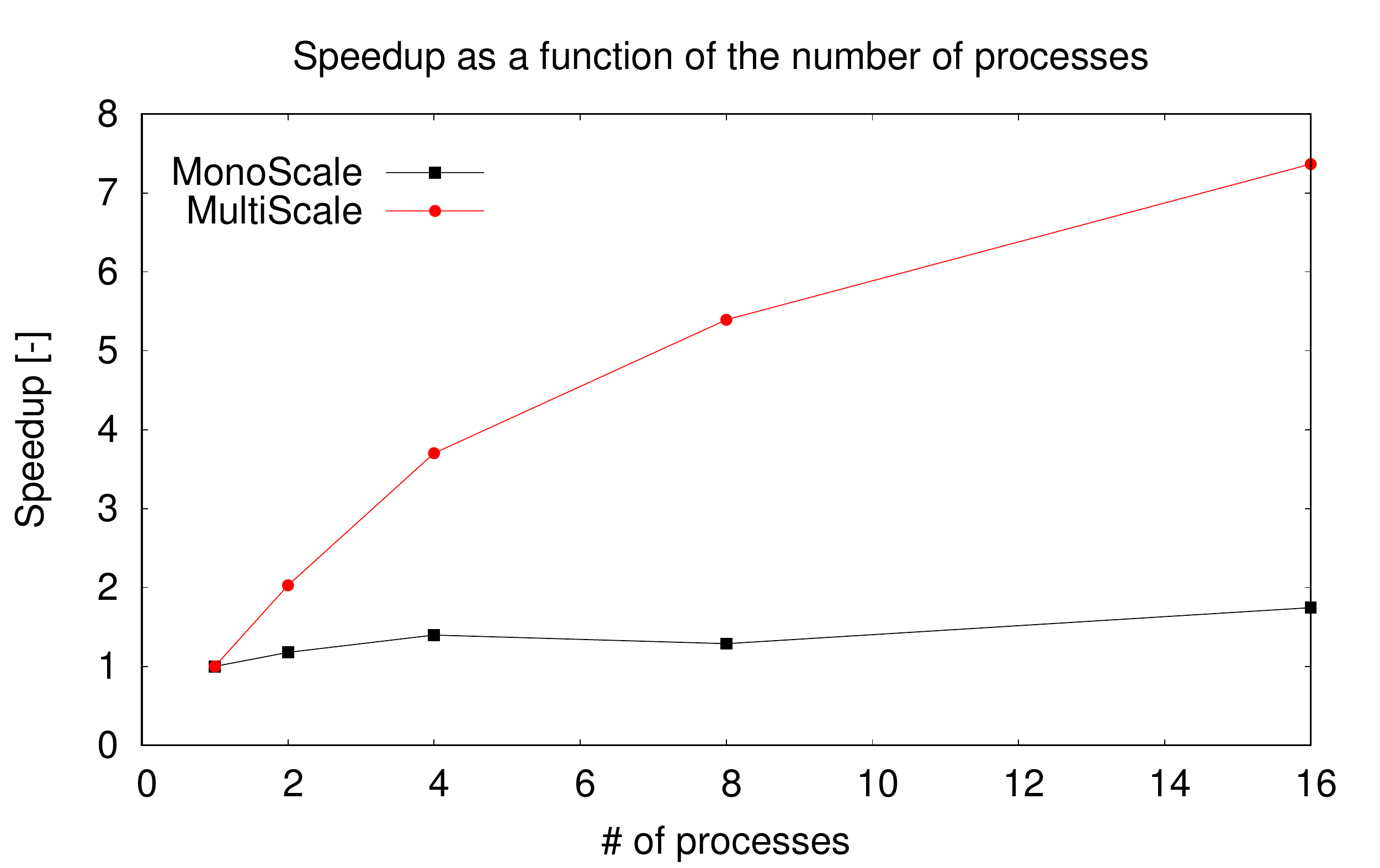}
 \caption{\label{DamBreakScal} Three-phase dam-break. Speedup as a function of the number of processes.
 better parallel performance is observed for the multiscale approach.}
\end{figure}
In this section, we propose the parallel execution of a benchmark for the multiscale DEM-VOF method.
The aim is of showing how, with non geometrically uniform cases, the higher flexibility
obtained by operating with two grid offers significant advantages in terms of parallel performance.
With this purpose, we compare the parallel execution of a mono-scale DEM-VOF coupling using a co-located partitions
strategy, with the parallel execution of a multiscale DEM-VOF coupling based on 
the parallelization strategy presented in this work.

We here re-propose the multiscale simulation of the three-phase dam break as in~\cite{PozzettiIJMF}.
As shown in figure
\ref{Dam_Break_Setup}, a configuration without 
intermediate obstacle is adopted, in 
a box of dimensions  $0.2 m$ x $0.1 m$ x $0.3 m$.
A column of water of extension $0.05 m$ x $0.1 m$ x $0.1 m$ with a uniformly 
layered bed of spherical particles at the bottom,
 is left breaking and stabilizing.
The spheres have a diameter of $2.7 \, mm$.
As already proposed in \cite{PozzettiIJMF}, we adopt a fine grid
discretized with $500k$ identical cubic 
cells while the coarse grid discretized with $6k$ identical volumes. 
For the mono-scale approach a domain discretization of $48k$ cells is 
chosen, and no coarse grid is used.
The simulation parameters are the standard ones as follows.
The liquid (heavy phase) density and viscosity are  taken as $1000 
kg/m^3$ and  $10^{-3}\, Pa \, s $ respectively,
while for the gas (light phase)  $1\, kg/m^3$ and  $10^{-5} \, Pa \, s $ are 
chosen.
The particle density is fixed at $2500 \,kg/m^3$. A linear dashpot impact model 
is used for both particle-particle and particle-container interactions
with spring constant of $1000 \, N/m$, a restitution coefficient of $ 0.9$ and 
a friction coefficient of $ 0.3$.

As already observed in figure~\ref{ParticleAcross}, figure~\ref{Fronts} confirms that the 
parallel execution with the strategy presented in this article, produces the same results as the sequential run 
and it is validated against experimental data. A further confirmation is provided in figure~\ref{Kin_en} that shows how 
also in the parallel execution, the kinetic energy reaches  proper convergence.

In figure~\ref{Dam_Break_Partitioning}, the partitioning for the two cases is showed. One can observe how,
in the mono-scale case, both the CFD and the DEM domains are partitioned uniformly in order to ensure 
perfect alignment between the partitions, according to a standard co-located partitions strategy.
In this case, due to the non-uniform distribution of the particles, this constraint results in all the particles to 
be assigned to a single process.
On the other hand, in the multiscale approach, the perfect alignment between the two domains is ensured by choosing
the coarse grid to be equal to the DEM grid. Therefore, the fine grid can 
be distributed in an arbitrary way, trying to optimize the ratio between the load balance and the communication.

In ~\cite{PozzettiIJMF} was discussed how, despite the usage of a finer CFD grid makes the multiscale approach more computationally expensive,
the overall computation time is comparable with the one involved in a mono-scale simulation for what concerns the 
sequential execution.
This is confirmed in figure~\ref{DamBreakTime}, where the computational time of the simulation time is depicted as a function of the number of processes for
both the mono-scale and the multiscale approaches.
One can observe how the sequential run of the multiscale simulation
requires more time.
Nevertheless, the enhanced flexibility offered for the parallel execution, allows it to perform better when operated over multiple processes.
In particular, when using more than 2 processes, the multiscale approach
is indeed faster than the mono-scale.
This can be explained by observing figure~\ref{DamBreakScal} where the speedup of the two approaches as a function of the number of processes
is proposed. It is clearly shown how the multiscale approach performs significantly better than the mono-scale one. 

In conclusion, for this non-uniform case, the grid-based parallelization strategy proposed for the multiscale approach
allows adopting a more efficient load distribution, leading to the counter-intuitive result of allowing a simulation that uses
an heavier CFD grid to run faster than one that uses a coarser one.
Therefore, within a co-located partitioning strategy, the multiscale approach is shown to provide not only
a better accuracy but also a reduced computational time when parallel execution is involved.

\subsection{10 Million Particles in 10 Million Cells}\label{10M10M}
\begin{figure}[ht]
\centering
 \includegraphics[width=0.6\textwidth]{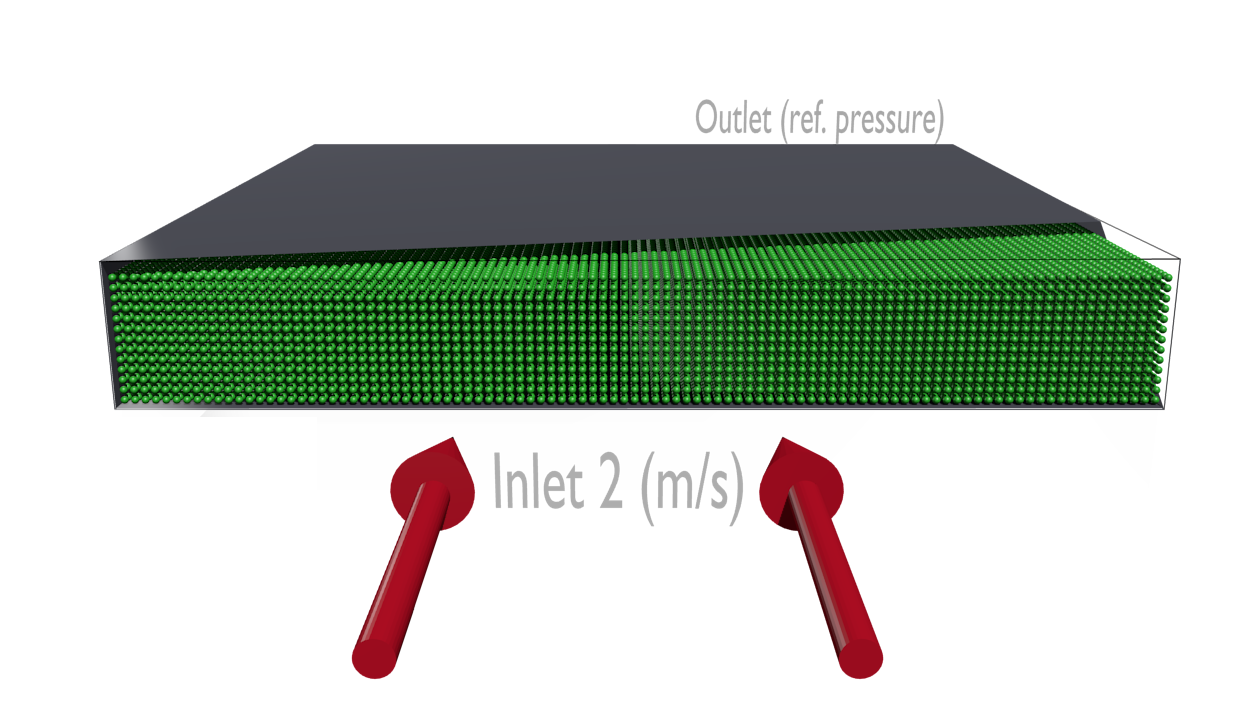}
 \caption{\label{10MParticlesSetup} 10 Million Particles in 10 Million cells. Simulation Setup.}
\end{figure}

\begin{figure}[ht]
\centering
 \includegraphics[width=0.6\textwidth]{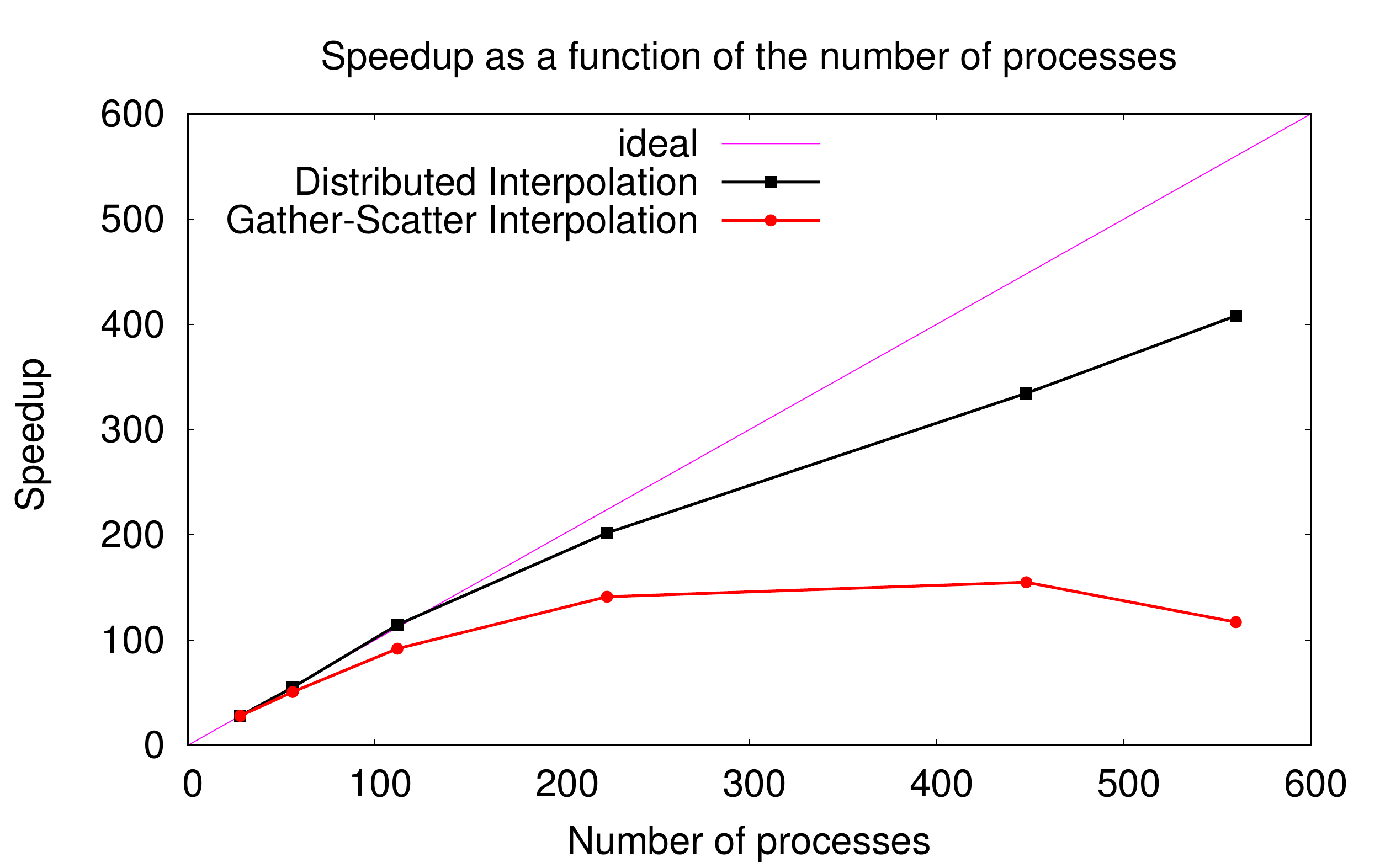}
 \caption{\label{ScalabilityGlobal} 10 Million Particles in 10 Million cells. Speedup as a function of time when using 
 gather-scatter and distributed communication for the grid-based interpolator: Significant influence of the communication
 strategy can be observed while running on more than 100 processes.}
\end{figure}

\begin{figure}[ht]
\centering
 \includegraphics[width=0.6\textwidth]{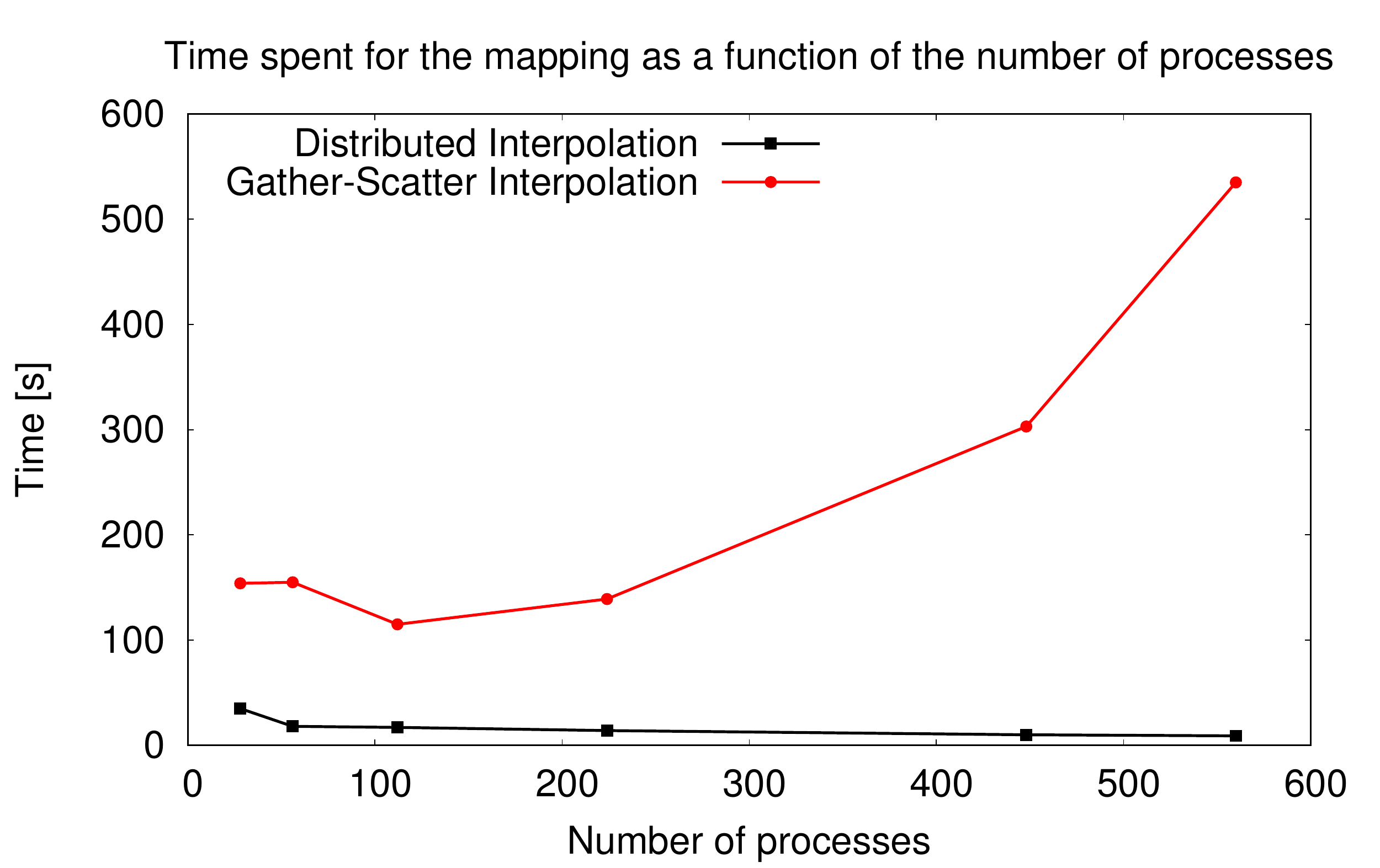}
 \caption{\label{TimeMapping} 10 Million Particles in 10 Million cells. Time spent in the interpolation between grid as a function of the
 number of processes when using gather-scatter and distributed communication for the grid-based interpolator:
 significant difference can be observed for the two communication strategies.}
\end{figure}
In this section, a case consisting of a layered bed of ten million particles
moving in the presence of a carrier gas is proposed. This case originally suggested in~\cite{SediFoam},
we here propose the results concerning the parallel execution of the benchmark with a dual-grid coupling 
in order to test the performance of the parallelization scheme when applied to a large scale problem.

As shown in figure~\ref{10MParticlesSetup}, the test case features a domain with sizes of 480mm x 40mm x 480 mm, as 
originally presented in~\cite{SediFoam}.
A layered bed of 10 million of particles is moving under the action of an incompressible flow.
The boundary conditions for the fluid solution are of constant Dirichlet at the inlet ($2 m/s$), 
non-slip at the wall boundaries and reference atmospheric pressure at the outlet.

In figure~\ref{ScalabilityGlobal}, the scalability performance of the coupled XDEM-OpenFOAM run is presented.
One can observe how, differently from what observed in \cite{SediFoam}, good scalability performance 
when operating over more than 500 processes is achieved. This represents a key improvement for the 
application of the method to large-scale problems. Furthermore,  it can be clearly noticed how the 
communication strategy chosen for the grid-based interpolator plays an important role in the performance of the 
overall code when an high number of processes is involved. In particular, while when using less than 100 processes,
the gather-scatter strategy seems not to penalize the execution excessively, when adopting 500 processes, 
the time spent in grid-communication becomes the main bottleneck of the simulation.

\begin{figure}[ht]
\centering
 \includegraphics[width=0.7\textwidth]{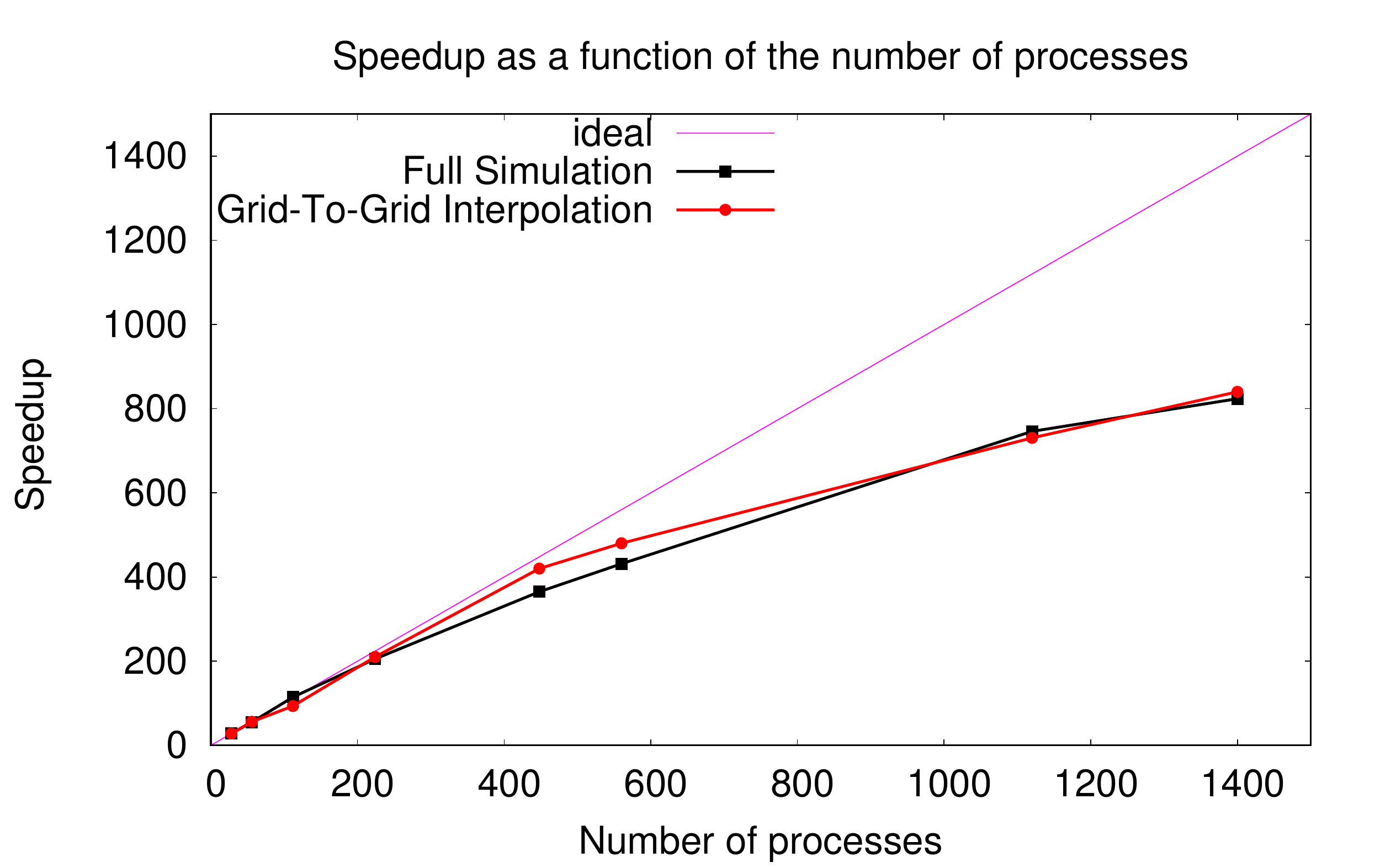}\qquad
 \caption{\label{ScalabilityLong}10 Million Particles in 10 Million cells. Speedup as a function of time of the full simulation
 and the grid-to-grid interpolation using a distributed communication for the grid-based interpolator: Good scalability properties of 
 both can be observed when running up to 1400 processes. }
\end{figure}

This is better shown in figure~\ref{TimeMapping}, where the time spent in the mapping operation during the first
50 times-steps of the simulation is depicted as a function of  the number of processes. Figure~\ref{TimeMapping} shows how 
a gather-scatter interpolation is always worse than a distributed one when operating on more than 28 processes. Furthermore,
when using more than 100 processes, the global time required by the interpolation increases with the number of processes when 
using a gather-scatter approach, while decreases when adopting a distributed one.
This shows how, for the current approach, a gather-scatter communication for the grid interpolation is only acceptable 
when a low number of processes are involved, while for approaching large-scale problems, a more sophisticated communication 
strategy must be chosen.

In figure~\ref{ScalabilityLong}, we report the parallel performance of the coupled XDEM-OpenFOAM run over 1400 processes.
Differently from figure~\ref{ScalabilityGlobal}, we here omit the results obtained with the gather-scatter approach as, when using more than
500 processes, the cost of this operation becomes prohibitive. 
For similar cost reasons, the scalability of the interpolation is obtained with reference to the 28 process run.
It can be observed how, when using a distributed approach for the 
interpolation, good scalability performance can be obtained for both the coupled run as well as the sole interpolation operation.
This represents an important results since it allows obtaining a speedup of $\sim 830$ that is approximatively 3.5 times better than what 
obtained by the previous literature for this test case~\cite{Pozzetti-co-located,SediFoam}.

\section{Conclusions}
A parallel dual-grid multiscale approach to CFD-DEM couplings has been presented.
It is based on the usage of two different CFD grids each associated with a different scale, and extends the
multiscale DEM-VOF method to general couplings for large-scale scenarios.
The innovative parallel implementation consists of adopting perfectly aligned co-located partitions between the DEM domain and the CFD domain associated 
with the bulk scale, while allowing independent discretization for the CFD domain associated with the fine scale.
This allows avoiding inter-process communication between the CFD and the DEM part and, at the same time, 
keeping flexibility on the domain partitioning.

The communication between the grid associated with the bulk scale and the one associated with the fine scale 
is operated through  a parallel grid-based interpolator.
The data structure that requires being communicated by this interpolator has been presented and two possible solutions for the 
parallel communication discussed.

Three benchmark cases have been discussed to assess consistency and performances of the proposed strategy.
Results showed how the higher flexibility, obtained by the independent partition of the fine mesh, allows the dual-grid multiscale approach to achieve
better parallel performance than a single-grid CFD-DEM coupling for inhomogeneous cases.
Furthermore, it was shown how the implementation of the parallel grid communicator plays a fundamental role in case 
of high computational loads. In particular, a gather-scatter communication strategy for the interpolator starts significantly affecting 
the performance of the overall numerical scheme when operated on more than 100 processes.

One of the main benefits of the current strategy consists of the possibility of adopting more complex partitioning algorithms
for the DEM domain and the CFD fine grid, that can, in general, be completely independent from one another.
Nevertheless, we want to point out how, using completely independent partitions for CFD and DEM increases the communication
costs of the interpolation between grids. Therefore, in order to obtain an optimal partitioning algorithm, this communication must be taken into account.
Finally, the definition of optimal partitioning algorithms and interpolations will represent an important topic for future studies.

\section*{Acknowledgments}
This research is in the framework of the project DigitalTwin, supported by the 
programme “Investissement pour la comp\'etitivit\'e et emploi” - European Regional Development Fund (Grant agreement: 2016-01-002-06).
The experiments presented in this paper were carried out
using the HPC facilities of the University of Luxembourg \cite{VBCG_HPCS14}.

\section*{References}

\bibliographystyle{plain}%

\end{document}